%% file: main.tex
\documentclass[lettersize,journal]{IEEEtran}
\usepackage{amsmath,amsfonts}
\usepackage{algorithmic}
\usepackage{algorithm}
\usepackage{array}
\usepackage{textcomp}
\usepackage{stfloats}
\usepackage{url}
\usepackage{verbatim}
\usepackage{graphicx}
\usepackage{cite}

\usepackage[pagebackref=true,breaklinks=true,letterpaper=true,colorlinks,bookmarks=false]{hyperref}
\usepackage{subcaption}
\usepackage{booktabs}
\usepackage{xcolor}
\newtheorem{assumption}{Assumption}

\begin{document}

\title{Multi-Disease-Aware Training Strategy for Cardiac MR Image Segmentation}

\author{Hong Zheng\textsuperscript{1,2,4}, Yucheng Chen\textsuperscript{3}, Nan Mu\textsuperscript{1,4,5}, and Xiaoning Li\textsuperscript{*,1,4,5}
\thanks{1. College of Computer Science, Sichuan Normal University, Chengdu, 610101, China}
\thanks{2. School of Computing and Artificial Intelligence, Southwest Jiaotong University, Chengdu, 611756, China}
\thanks{3. Department of Cardiology, West China Hospital, Sichuan University, Chengdu, 610041, China}
\thanks{4. Visual Computing and Virtual Reality Key Laboratory of Sichuan Province, Chengdu, 610066, China}
\thanks{5. Sichuan 2011 Collaborative Innovation Center for Educational Big Data, Chengdu, 610066, China}
\thanks{*Corresponding author: lxn@sicnu.edu.cn}
}



\maketitle

\begin{abstract}
Accurate segmentation of the ventricles from cardiac magnetic resonance images (CMRIs) is crucial for enhancing the diagnosis and analysis of heart conditions. Deep learning-based segmentation methods have recently garnered significant attention due to their impressive performance. However, these segmentation methods are typically good at partitioning regularly shaped organs, such as the left ventricle (LV) and the myocardium (MYO), whereas they perform poorly on irregularly shaped organs, such as the right ventricle (RV). In this study, we argue that this limitation of segmentation models stems from their insufficient generalization ability to address the distribution shift of segmentation targets across slices, cardiac phases, and disease conditions. To overcome this issue, we present a Multi-Disease-Aware Training Strategy (MTS) and restructure the introduced CMRI datasets into multi-disease datasets. Additionally, we propose a specialized data processing technique for preprocessing input images to support the MTS. To validate the effectiveness of our method, we performed control group experiments and cross-validation tests. The experimental results show that (1) network models trained using our proposed strategy achieved superior segmentation performance, particularly in RV segmentation, and (2) these networks exhibited robust performance even when applied to data from unknown diseases. 
\end{abstract}

\section{Introduction}
\label{sec1}
\IEEEPARstart{A}{ccurate} diagnosis of heart conditions and appropriate therapeutic schedules are crucial, given that cardiovascular diseases (CVDs) stand as the leading cause of global mortality, claiming an estimated 17.9 million lives annually, according to the World Health Organization (WHO). Artificial intelligence (AI)-aided diagnosis methods have garnered significant research attention due to their high accuracy, fast prediction capabilities, and low human cost. Within this realm, the precise delineation of the heart's structures from cardiac magnetic resonance images (CMRIs) emerges as a pivotal stage. Over recent decades, the advancement of convolutional neural network (CNN) segmentation models has been remarkable, not only enhancing segmentation accuracy but also propelling the advancement of computer vision (CV) techniques. However, CNN-based segmentation models (segmenters) are typically good at partitioning regularly shaped organs, such as the left ventricle (LV) and the myocardium (MYO), whereas they perform poorly on irregularly shaped organs, such as the right ventricle (RV). This limitation of CNN-based segmentation models has been demonstrated in studies~\cite{b2,b3,b8}, and we attribute it to the limitations of risk-minimization-based training strategies. 

\begin{figure}[t]
\centering
\includegraphics[width=0.85\linewidth]{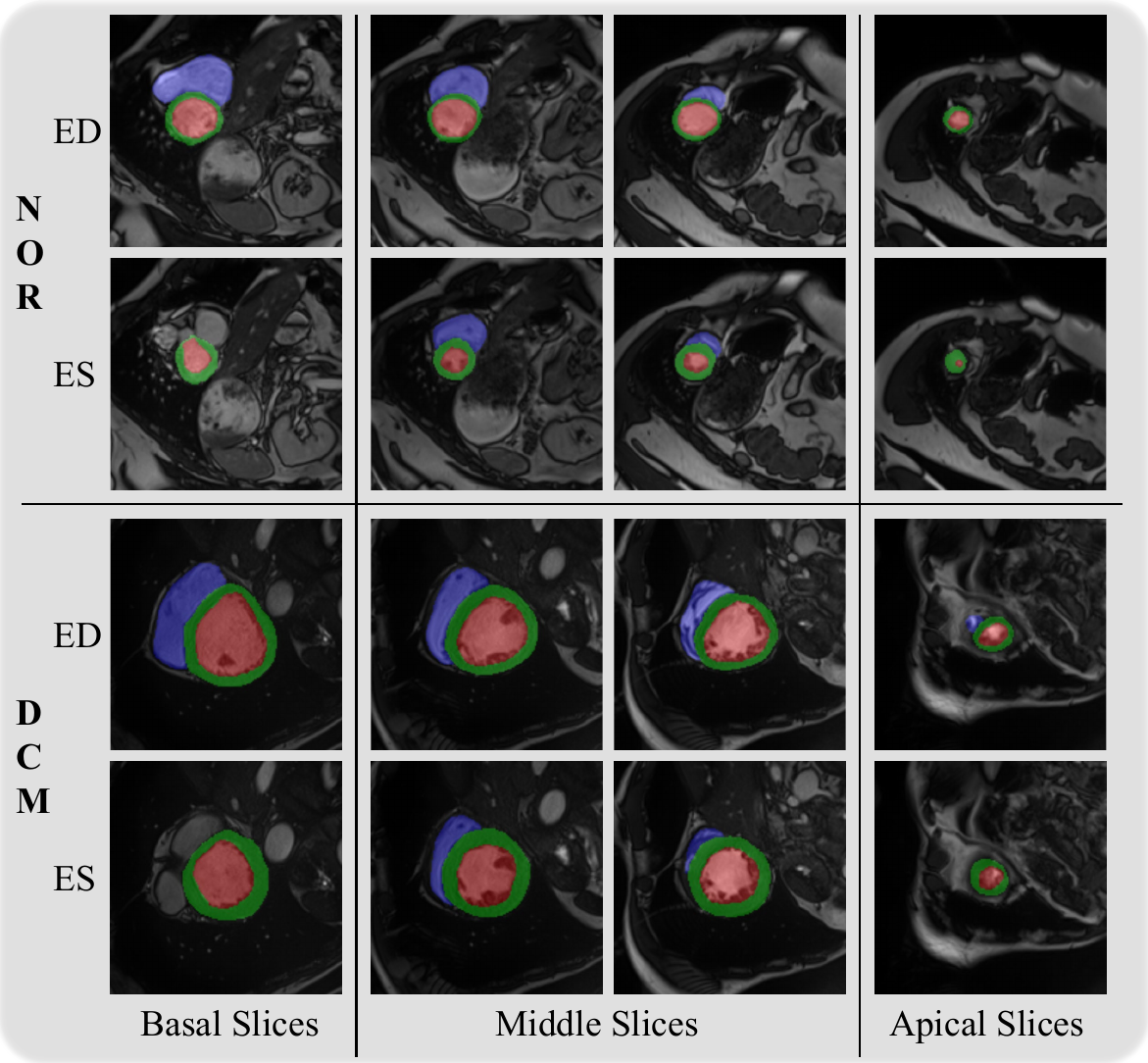}
\caption{Several examples of cardiac images from different slices. NOR and DCM represent normal subjects and dilated cardiomyopathy, and ED and ES denote end-diastolic and end-systolic phases. The blue, green, and red parts in the figures are the RV, MYO, and LV, respectively.}
\label{fig1}
\end{figure}

For a more thorough exploration of the aforementioned limitation, we propose observing two cardiac images depicting different phases and slices across distinct diseases (see Figure~\ref{fig1}). Upon observation, it becomes evident that the RV (blue area) undergoes more pronounced shape changes across various phases (ED and ES) and slices (apical to basal slices) compared to other cardiac organs. Notably, this shape change of the RV also appears in different diseases (NOR and DCM). Namely, we considered that the distribution of the RV differs significantly from that of the LV and MYO in CMRIs. Consequently, during network training, the gradient optimization direction was adjusted to minimize the risk loss by prioritizing the more easily distinguishable distributions of the LV and MYO. While this learning strategy yields an optimal validation loss, it may result in suboptimal segmentation performance for the RV. In conclusion, risk-minimization-based training strategies are limited in their ability to address the distribution shift of segmentation targets across slices, cardiac phases, and disease conditions. 

Recently, there has been increased attention on domain generalization~\cite{b42} (DG) studies, focusing on addressing the domain shift problem. Several studies, such as~\cite{b36, b37}, aim to learn a mapping that performs well across a large set of related yet unobserved domains by using a dataset that includes multiple observed training domains. Due to the differing distributions across domains, a common training objective known as min-max minimization is introduced to guide network model training~\cite{b36}. This training objective aims to minimize the maximum difference among various domain distributions, thereby mitigating the domain shift problem and enhancing the generalization ability of network models. It suggests we can treat the distribution shift problem of segmentation targets as a domain shift problem, thereby designing a training strategy based on min-max minimization. 

In this study, we refer to the proposed training method as the Multi-Disease-Aware Training Strategy (MTS). To perform this training strategy, we restructure the training sets of the introduced CMRI datasets (two short-axis CMRI datasets, one public and one private) into multiple disease-specific datasets. Additionally, we propose a specialized data processing technique for preprocessing input images to support MTS. This method is not a typical augmentation technique; instead, it serves as a data “destroyer” aimed at erasing parts of the information in the input image data during the training stage. This is because we found that the segmentation performance of trained segmenters is better when using the destroyed data for training, and we refer to the specially processed data as Incomplete Training Data (ITD). By incorporating ITD into MTS, a new training strategy called MTS plus ITD (MTS+ITD) is established. Through control group experiments, we found that the segmentation accuracy on all introduced datasets from segmenters trained by MTS+ITD increased, particularly in RV segmentation. Furthermore, we conducted cross-validation tests where network models trained on one (A) dataset were used to segment another (B) dataset, evaluating the generalization ability of models in processing unknown disease data. All of our experimental results strongly prove the feasibility and effectiveness of the proposed methods. 

In conclusion, our work innovates by considering the distribution shift problem of segmentation targets as a domain shift problem. We make two key contributions: \textbf{firstly}, we demonstrate the impact of different training strategies on the segmentation and generalization abilities of network models, and \textbf{secondly}, we reveal the advantages of incomplete data over complete data for training networks.

\section{Related Works}
\label{sec2}
\subsection{Normal Training Strategies}
Segmentation networks are typically trained using a universal objective, such as cross-entropy or mean squared error, to minimize validation risk and perform well on testing data. We refer to these training strategies as the normal training strategy (NTS). For example, some noteworthy works include those by P.V. Tran~\cite{b2}, M. Khened et al.~\cite{b12}, R.P. Poudel et al.~\cite{b13}, Chakravarty et al.~\cite{b14}, Ammar et al.~\cite{b21}, etc. Despite using the NTS, these studies typically focused on enhancing segmentation models based on the fully convolutional network~\cite{b1} (FCN). The UNet architecture~\cite{b3} serves as a prime example, with various segmentation studies adopting it as their backbone models, including works by {\"O.} {\c{C}}i{\c{c}}ek et al.~\cite{b4}, Huang et al.~\cite{b20}, H. Zheng et al.~\cite{b8}, J. Schlemper et al.~\cite{b15}, F. Isensee et al.~\cite{b27}, and others. Additionally, many researchers have integrated Attention mechanisms~\cite{b9} into segmentation networks to improve feature extraction from input images. Examples of such methods include CBAM~\cite{b10}, spatial-temporal attention models~\cite{b11}, attention gate~\cite{b15}, multi-scale self-guided attention~\cite{b16}, and transformer-based T-AutoML~\cite{b31}. 

\subsection{Priors-based Training Strategies}
Several approaches exist for incorporating prior knowledge into NTS, with their training methods considered as Prior-based Training Strategies (PTS). For instance, O. Oktay et al.~\cite{b6} and C. Chen et al.~\cite{b7} introduced training methods that learn anatomical shape priors across different 2D standard views (short-axis and long-axis CMRI) and use them to enhance segmentation accuracy. H. Zheng et al.~\cite{b8} proposed the Cross U-net, which also integrates reconstructed image features as prior knowledge to guide the segmentation task. Furthermore, the benefits of utilizing shape prior knowledge have been observed in the works of C. Zotti et al.~\cite{b17}, Q. Zheng et al.~\cite{b18}, and Q. Yue et al.~\cite{b19}. More recently, K. Zhang et al.~\cite{b25} and A. Raju et al.~\cite{b35} presented holistic strategies and deep implicit statistical shape models. Additionally, some researchers have presented similar PTS methods for their segmentation tasks, including X. Chen et al.~\cite{b24}, C. Seibold et al.~\cite{b29}, J. Wang et al.~\cite{b30}, X. Yang et al.~\cite{b23}, H. Tang et al.~\cite{b28}, among others.

\subsection{Multi-domain Training Strategies}
Training network models using multi-domain datasets is known as Multi-domain Training Strategies (MDTS). Domain adaptation (DA)~\cite{b41} and domain generalization (DG)~\cite{b42} typically involve employing MDTS to train robust models, such as some learning theories like IRM~\cite{b36}, P-IRM~\cite{b37}, and IIB~\cite{b38,b39}. DA and DG primarily tackle the domain shift problem under conditions that include multiple data distributions, as the assumption of identically and independently distributed data is often unrealistic for the practical deployment of machine learning models. Further discussion on this topic is beyond the scope of this study, but interested readers can refer to \cite{b42}. With the conclusion of the M\&Ms challenge~\cite{b22}, research on domain generalization in medical image processing has garnered significant attention, as evidenced by the works of C. Huang et al.~\cite{b20} and H. Yao et al.~\cite{b26}. Additionally, cross-dataset collaborative learning~\cite{b32} and multi-modal learning~\cite{b33,b34,b40} are also widely studied.

\section{Methodologies}
\label{sec3}
This section will present our proposed multi-disease-aware training strategy (MTS) and data preprocessing method. First, we present the following hypothesis as the foundation of our work: 
\begin{assumption}
\label{assu1}
For cardiac MRI multi-target segmentation problems, we assume that poor segmentations produced by segmentation models are causally linked to the distribution shift of segmentation targets across different slices, cardiac phases, and disease conditions. 
\end{assumption}

Based on this assumption and studies on domain generalization (DG), we conclude that enhancing the generalization capability of segmentation models across different slices, phases, and diseases is equivalent to improving segmentation performance for any target organ in CMRIs. To enhance the generalization capability of segmenters, we restructure the training datasets into multiple disease-specific datasets (see Section~\ref{sec3a}), develop a data preprocessing method specifically for the input training data (see Section~\ref{sec3b}), and formulate a learning algorithm (see Section~\ref{sec3c}).

\subsection{Multi-disease Dataset \& Learning objective}
\label{sec3a}
This subsection firstly demonstrates the process of reconstructing the training datasets from the introduced CMRI datasets into multi-disease datasets using a {descriptor-based} approach. We assume that a dataset ${D_k}: = \{ (X_i^k,Y_i^k)\} _{i = 1}^{{n_k}}$ contains ${n_k}$ data samples for a single disease $k \in K_{tr}$, where $K_{tr}$ represents the set of all diseases for training, with a total of $\left| K_{tr} \right| $ diseases. The symbol $K$ originates from the German word \textit{Die Krankheit} meaning “the disease”. The pair $(X^k, Y^k)$ denotes a CMRI and its corresponding ground-truth (GT) segmentation map, where $ {X^k} \in {{\mathbb R}^{P \times H \times W \times Z}}$ and $ {Y^k} \in {{\mathbb R}^{P \times H \times W \times Z}}$. Here, $P$, $H$, $W$, and $Z$ represent the phase number, height, width, and the number of slices along the $z$-axis, respectively. 

Based on the definition of ${D_k}$ above, we provide an example for further illustration. Consider an original training dataset consisting of five disease types, each with fifty cases (data samples). We can then easily establish the set $ K_{tr}$ with $\left| K_{tr} \right| = 5$, and the set ${D_k}$ with ${n_k} = 50$. Note that $n_{k_1} \ne n_{k_2},\forall k_1, k_2 \in K_{tr}$ for most practical scenarios. Here, ${n_k}$ denotes the number of cases, where each case corresponds to a 4D cardiac MRI (CMRI) matrix. This 4D matrix is typically divided into two phases: the end-diastolic (ED) and end-systolic (ES) phases and the remaining phases are not valuable for observation. Each of the ED and ES phases is represented as a 3D matrix, which is then dimensionally reduced along the $z$-axis to yield a 2D matrix. For example, consider a 4D matrix with dimensions $(12,128,128,9)$. A 3D matrix in either ED or ES phase has dimensions $(128, 128, 9)$. Consequently, we can derive nine 2D matrices, each with dimensions $(128, 128)$. Hereafter, the dataset ${D_k}$ with $n_k$ samples denotes the 2D image data along with their GT segmentation maps, i.e., $ {X^k} \in {{\mathbb R}^{H \times W}}$, $ {Y^k} \in {{\mathbb R}^{H \times W}} $ and $n_k= 2 \times Z \times n_k $.

To enhance comprehension of our motivations, we address the following questions. (1) Why do we divide the training datasets by the type of disease instead of others? In Figure~\ref{fig1}, we observed that different diseases exhibited vastly distinct RV structures. Therefore, we argue that the standard for division should be the disease rather than the slice or the phase. (2) How do we handle the distribution differences in RV exhibited in different slices and phases? We employ the random-sampling technique to randomly select data from $D_k, \forall k \in K_{tr}$ and then construct the sampled data pair $(X^k, Y^k)$ into an input subset ${\bf{d}}: = \{ ({X^k},{Y^k})\} _{k = 1}^{\left| {{K_{tr}}} \right|}$ during the training stage. Note that this subset ${\bf{d}}$ contains $ \left| {{K_{tr}}} \right|$ random slice and phase data pairs. For convenience, we omit the batch size number. If considering it, the subset can be denoted as $ {{\bf{d}}_b}: = \{ (X_b^k,Y_b^k)\} _{k = 1}^{\left| {{K_{tr}}} \right|},(X_b^k,Y_b^k) \subset \{ (X_i^k,Y_i^k)\} _{i = 1}^{{n_k}}$, where $b$ is the batch size. However, hereafter, we neglect this. 

Since ${\bf{d}}$ contains different slices, phases, and diseases, the goal of minimizing distribution differences in those has been established. Correspondingly, we formulate a min-max training objective as follows: 
\begin{equation}
\label{eq1}
\mathop {\min }\limits_f \sum\limits_{k \in {K_{tr}}} {{R^k}(f)},
\end{equation}
where $ {R^k}(f): = {{\mathbb E}_{{X^k},{Y^k}}}[\ell (f({X^k}),{Y^k})]$, $ f:X \to \hat Y $ is a segmenter and is parameterized, and $\ell ( \cdot )$ normally denotes a loss function. The predicted segmentation results are denoted as ${{\hat Y}^k} = f({X^k})$, where ${{\hat Y}^k} \in {{\mathbb R}^{H \times W}}$. 

\begin{algorithm}[t]
\caption{Multi-disease-aware Training Strategy (MTS)}
\label{alg1}
\textbf{Input}: $ \{ {D_k}|\forall k \in {K_{tr}}\} $.\\
\textbf{Parameter}: Epoch, KEY, BatchSize $b$.\\
\textbf{Output}: $f$.
\begin{algorithmic}[1] 
\WHILE{Epoch}
\WHILE{$\forall {{\bf{d}}_b} \in \{ {D_k}|\forall k \in {K_{tr}}\}$ and $ \max ({n_k}/b)$}

\WHILE{$\forall (X_b^k,Y_b^k) \in {{\bf{d}}_b}$}

\IF{KEY}
\STATE $X_b^k = X_b^k \cdot {H_{ideal}}$.
\ENDIF

\STATE Solve for $\hat Y_b^k$ using $f(X_b^k)$.
\STATE Solve for $\sum\nolimits_{k \in {K_{tr}}} {{R^k}(f)} $ using $ Y_b^k $ and $\hat Y_b^k$. 
\ENDWHILE

\STATE Optimize $f$ using $\sum\nolimits_{k \in {K_{tr}}} {{R^k}(f)} $.

\ENDWHILE
\ENDWHILE
\end{algorithmic}
\textbf{Note}: The maximum time complexity is $O(n^3)$ on the GPU time.
\end{algorithm}

\subsection{Data Preprocessing Method}
\label{sec3b}
This subsection will detail how to preprocess the input data during the training stage. As previously mentioned, this method is not a technique for data augmentation; instead, it aims to destroy some information in the input images. To achieve this, we use the rectangular ideal mask as the primary method for selecting and excluding information from the original input data. The rectangle ideal mask (see Figure~\ref{fig7}) can be formulated as follows:
\begin{equation}
\label{eq2}
{H_{ideal}}(u,v) = \left\{ {\begin{array}{*{20}{c}}
{0,}&{u\& v \in Box({u_c},{v_c},\alpha )}\\
{1,}&{else}
\end{array}} \right.,
\end{equation}
where $ u \in \{ 0, \ldots H - 1\} $ and $v \in \{ 0, \ldots W - 1\} $. $Box({u_c},{v_c},\alpha )$ represents a rectangular box with its center at coordinates $({u_c},{v_c})$ and sides of length $\alpha$. We empirically set $\alpha$ as $\alpha  = \lambda  \cdot \min (H,W)$. It is important to note that the center $({u_c},{v_c})$ is not fixed; rather, it undergoes random changes. Specifically, $ {u_c} \in \{ \alpha /2, \ldots (H - \alpha )/2\} $ and $ {v_c} \in \{ \alpha /2, \ldots (W - \alpha )/2\}$. 

Applying ${H_{ideal}}(u,v)$ to the original data, we can generate new input data. For instance, consider a CMRI $X$ as the original data. We then multiply $X$ by ${H_{ideal}}(u,v)$ to obtain a new image, denoted as ${X_{ideal}} = X \cdot {H_{ideal}}$. Subsequently, $X$ is replaced by ${X_{ideal}}$ as the input data.  

Several examples of processed $X_{ideal}$ are listed in Figure~\ref{fig8}, where it is evident that some information in $X_{ideal}$ is lost. Due to random sampling, the missing part could appear anywhere within $X_{ideal}$. Consequently, networks cannot process the same image data twice during training, resulting in a latent increase in the number of data samples. For instance, if $n_k = 100$ and $ Epoch = 10$, the total number of data samples for a single disease $k$ is $ 10 \times 100 = 1000 $. Importantly, only the training data undergo processing, while the testing data remains in its original form.

\subsection{MTS Algorithm}
\label{sec3c}
The overall algorithm of MTS is shown in Algorithm~\ref{alg1}. The input is $ \{ {D_k}|\forall k \in {K_{tr}}\} $, with parameters Epoch, KEY, and Batch Size: $b$, and the output is the predictor $f$. Epoch defines the maximum training period. The parameter KEY controls whether the input training data should be preprocessed. If KEY is true, we preprocess the input data, and the training strategy is MTS plus incomplete training data (MTS+ITD); if KEY is false, the input data keep their original formations, and the training strategy is MTS plus complete training data (MTS+CTD). As previously mentioned, $ {{\bf{d}}_b}$ is a random-sampled subset of $ \{ {D_k}|\forall k \in {K_{tr}}\} $. 

Algorithm~\ref{alg1} introduces a key departure from the normal training strategy outlined in Algorithm~\ref{alg2}. Rather than randomly selecting image-GT label pairs from an entire dataset $D$ that do not distinguish between specific diseases, MTS requires that ${{\bf{d}}_b}$ must contain all kinds of diseases, i.e., distinguishing them. The inclusion of different phases and slices for ${{\bf{d}}_b}$ is not 100\% guaranteed due to the random sampling method. However, if the EPOCH is large, we can approach the probability to 1.0.

\section{Experiments}
\label{sec4}

\subsection{Datasets and Others}
\label{sec4a}
\textbf{Datasets}: The experiments involved two cardiac short-axis cine-MRI datasets: one sourced from the Automated Cardiac Diagnosis Challenge (ACDC)~\cite{b5}, and the other obtained from the Department of Cardiology, West China Hospital, Sichuan University (DCWC). The ACDC dataset used in this study comprises 150 patients from the targeted population, categorized into five disease groups: normal subjects (NOR), previous myocardial infarction (MINF), dilated cardiomyopathy (DCM), hypertrophic cardiomyopathy (HCM), and abnormal right ventricle (ARV). This indicates that $ \left| {{K_{tr}}} \right| =  5$. The ACDC dataset is further divided into training (100 cases) and testing (50 cases) groups. In the training group, there are 20 cases of NOR, 20 cases of MINF, 20 cases of DCM, 20 cases of HCM, and 20 cases of ARV. Meanwhile, in the testing group, NOR, MINF, DCM, HCM, and ARV contain 10 cases each.

Similarly, the DCWC dataset used in this study consists of 150 patients from the targeted population, classified into four disease groups: normal subjects (NOR), dilated cardiomyopathy (DCM), hypertrophic cardiomyopathy (HCM), and pulmonary artery hypertension (PAH). This indicates that $ \left| {{K_{tr}}} \right| = 4$. The DCWC dataset is further divided into training (100 cases) and testing (50 cases) groups. In the training group, there are 40 cases of NOR, 20 cases of DCM, 20 cases of HCM, and 20 cases of PAH. Meanwhile, in the testing group, NOR contains 20 cases, and DCM, HCM, and PAH contain 10 cases each. 

The ACDC dataset is publicly available, whereas the DCWC dataset is privately sourced. The DCWC data was generated with the assistance of experts from West China Hospital. The labeling technique involves cross-dataset segmentation using a UNet model trained on the ACDC data, with subsequent refinement by experts. Consequently, it is inevitable for the DCWC dataset to contain certain errors in the GT labels. We employed the DCWC to assess the robustness of our method when applied to short-axis cardiac image data from diverse source domains. 

For the 2D segmentation tasks, we resized all training image data samples including ACDC data and DCWC data to 2D image matrices with dimensions of $160 \times 160$, using cropping or padding techniques. Whereas, the testing data is kept in their original formations. 

\begin{algorithm}[t]
\caption{Normal Training Strategy (NTS)}
\label{alg2}
\textbf{Input}: $D: = \{ ({X_i},{Y_i})\} _{i = 1}^L$.\\
\textbf{Parameter}: Epoch, KEY, BatchSize $b$.\\
\textbf{Output}: $f$.
\begin{algorithmic}[1] 
\WHILE{Epoch}
\WHILE{$\forall (X_b, Y_b) \in {D}$ and $L / b$}

\IF{KEY}
\STATE $X_b = X_b  \cdot {H_{ideal}}$
\ENDIF

\STATE Solve for ${\hat Y_b}$ using $ f( X_b )$.
\STATE Solve for $\ell (\hat Y_b, Y_b)$ using $\hat Y_b$ and $Y_b$.

\STATE Optimize $f$ using $\ell$.

\ENDWHILE
\ENDWHILE
\end{algorithmic}
\textbf{Note}: The maximum time complexity is $O(n^2)$ on the GPU time.
\end{algorithm}

\textbf{Normal Training Strategy}: The normal training strategy (NTS) refers to combining all trainable data samples into a large dataset $ D: = \{ ({X_i},{Y_i})\} _{i = 1}^L$ without distinguishing between specific diseases, where $L$ denotes the number of the entire training sample $ (X,Y)$. Correspondingly, the training objective is 
\begin{equation}
\label{eq3}
\mathop {\min }\limits_f \ell (f(X),Y),
\end{equation}
where $f$ is parameterized and $\ell ( \cdot )$ denotes a loss function. Comparing to MTS, the training algorithm of NTS is exhibited in Algorithm~\ref{alg2}. Notably, there is one difference between traditional training plans, which is the parameter KEY. The parameter KEY controls whether the input training data is complete or not. If KEY is true, we have the training strategy: NTS plus incomplete training data (NTS+ITD); if KEY is false, we have the training strategy: NTS plus complete training data (NTS+CTD).

\textbf{Networks}: In this study, we only conducted two basal segmentation networks that are UNet and Transformer-based UNet (TUNet). UNet is the most popular network for segmentation tasks, and the Transformer module recently became a hot topic and is utilized by lots of studies. The architectures of two networks are $UNet = De(En(\cdot))$ and $TUNet = De(Trans(En(\cdot)))$, where $En(\cdot)$, $De(\cdot)$, and $Trans(\cdot)$ represent the encoder, decoder, and transformer module, respectively. Notably, the segmentation results of these two networks may not match state-of-the-art (SOTA) evaluations. This is because we did not elaborate on designing and modifying their structure to achieve the best segmentation accuracy. We only discuss the causality between different training strategies and segmentation results instead of the network itself. 

\textbf{Metrics}: We utilized the Dice score and Hausdorff distance (HD) to assess the accuracy of segmentation. The Dice metric measures the overlap between the predicted segmentation A and the GT segmentation B. It is defined as:
\begin{equation}
\label{eq4}
D{\rm{ice}} = \frac{{2\left| {A \cap B} \right|}}{{\left| A \right| + \left| B \right|}}.
\end{equation}
This metric yields a value between 0 and 1, where 0 indicates no overlap and 1 signifies perfect agreement. A higher Dice metric signifies better agreement.

The HD measures the maximum distances between the predicted segmentation contours $\partial A$ and the GT segmentation contours $\partial B$. It is defined as:
\begin{equation}
\label{eq5}
HD = \max [\mathop {\max }\limits_{p \in \partial A} d(p,\partial B),\mathop {\max }\limits_{q \in \partial B} d(q,\partial A)],
\end{equation}
where $d(p,\partial)$ represents the minimum distance from point $p$ to the contour $\partial$. A lower distance metric indicates better agreement.

\textbf{Training Details}: The experiment utilized a single Nvidia RTX 4090 24 GB GPU and the Tensorflow-gpu 2.6.0 deep learning library. To maintain consistency, other experimental settings, including the number of epochs (\textbf{50}), optimizer (\textbf{Adam}), learning rate (\textbf{0.001}), batch size (\textbf{8}), and loss function (\textbf{cross-entropy}), remained unchanged for all networks. Notably, we conducted \textbf{ten training runs} at least for each network to ensure the reliability and robustness of the experimental findings.

\subsection{Experiments on ACDC data}
\label{sec4b}

\begin{figure}[t]
\centering
\includegraphics[width=0.50\linewidth]{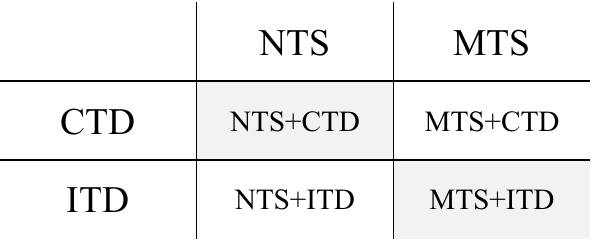}
\caption{A control group experimental illustration. The diagonal comparison is the main experiment, and the others belong to the ablation study. }
\label{fig2}
\end{figure}

\textbf{Experimental outline}: Based on two different training strategies and two different training data processing methods, we can design control group experiments (see Figure~\ref{fig2}). (1) Using NTS vs. using MTS when the training data is CTD (NTS+CTD vs. MTS+CTD); (2) Using NTS vs. using MTS when the training data is ITD (NTS+ITD vs. MTS+ITD); (3) Using CTD vs. using ITD when the training strategy is NTS (NTS+CTD vs. NTS+ITD); (4) Using CTD vs. using ITD when the training strategy is MTS (MTS+CTD vs. MTS+ITD); (5) NTS+CTD vs. MTS+ITD.

\begin{table*}[t]
\setlength{\abovecaptionskip}{0pt}
\setlength{\belowcaptionskip}{0pt}
\caption{Segmentation evaluations on the ACDC testing dataset, with results presented as the mean and standard deviation (Std) of various metrics, including Dice Scores (\%) and Hausdorff Distance (HD) in millimeters (mm). Bold highlights indicate improvements, while red indicates the best results. }
\centering
	\begin{subtable}{\linewidth}
	\centering
	\begin{tabular}{llrrrrrr}
	\toprule
	Training Strategies&Models&\multicolumn{2}{c}{LV Dice}&\multicolumn{2}{c}{RV Dice}&\multicolumn{2}{c}{MYO Dice}\\
	{}&{}&\multicolumn{1}{c}{ED}&\multicolumn{1}{c}{ES}&\multicolumn{1}{c}{ED}&\multicolumn{1}{c}{ES}&\multicolumn{1}{c}{ED}&\multicolumn{1}{c}{ES}\\
	\midrule
	NTS+CTD&UNet&96.3 (0.2)&91.5 (0.5)&92.9 (0.4)&86.5 (0.6)&88.9 (0.4)&90.4 (0.3)\\
	{     }&TUNet&96.5 (0.2)&91.9 (0.6)&93.0 (0.4)&87.4 (0.5)&\textcolor{red}{89.7} (0.3)&90.6 (0.3)\\
	MTS+ITD&UNet&\textcolor{red}{\textbf{96.6}} (0.1)&\textcolor{red}{\textbf{91.9}} (0.5)&\textcolor{red}{\textbf{93.7}} (0.4)&\textbf{88.4} (0.6)&{\textbf{89.6}} (0.3)&\textcolor{red}{\textbf{91.1}} (0.3)\\
	{     }&TUNet&96.5 (0.1)&91.9 (0.8)&\textbf{93.5} (0.3)&\textcolor{red}{\textbf{88.5}} (0.6)&89.5 (0.2)&\textbf{91.0} (0.3)\\
	\bottomrule
	\end{tabular}
	\caption{}
	\end{subtable}
	\begin{subtable}{\linewidth}
	\centering
	\begin{tabular}{llrrrrrr}
	\toprule
	Training Strategies&Models&\multicolumn{2}{c}{LV HD}&\multicolumn{2}{c}{RV HD}&\multicolumn{2}{c}{MYO HD}\\
	{}&{}&\multicolumn{1}{c}{ED}&\multicolumn{1}{c}{ES}&\multicolumn{1}{c}{ED}&\multicolumn{1}{c}{ES}&\multicolumn{1}{c}{ED}&\multicolumn{1}{c}{ES}\\
	\midrule
	NTS+CTD&UNet&4.3 (0.5)&7.4 (2.1)&12.7 (1.8)&12.9 (2.0)&6.0 (1.3)&7.4 (1.2)\\
	{     }&TUNet&5.4 (1.2)&8.4 (2.3)&12.2 (3.7)&12.2 (2.9)&7.0 (1.8)&10.1 (3.4)\\
	MTS+ITD&UNet&\textcolor{red}{\textbf{3.9}} (0.7)&\textcolor{red}{\textbf{4.6}} (1.2)&\textbf{6.8} (1.1)&\textbf{7.1} (0.6)&\textcolor{red}{\textbf{4.5}} (0.5)&\textcolor{red}{\textbf{5.2}} (0.8)\\
	{      }&TUNet&\textbf{4.6} (1.0)&\textbf{6.1} (1.3)&\textcolor{red}{\textbf{6.6}} (0.9)&\textcolor{red}{\textbf{6.9}} (0.8)&\textbf{5.3} (0.7)&\textbf{6.6} (1.6)\\
	\bottomrule
	\end{tabular}
	\caption{}
	\end{subtable}
\label{tab1}
\end{table*}

\textbf{Main segmentation results}: This experiment compares the results when using NTS and CTD training networks and the results when using MTS and ITD training networks. It is the diagonal comparison in Figure~\ref{fig2}, which is NTS+CTD vs. MTS+ITD. The segmentation accuracy of processing ACDC data is shown in Table~\ref{tab1}. According to this table, we can conclude two conclusions. (1) Different training strategies affect the segmentation ability of networks; (2) The segmentation ability of networks is enhanced by MTS+ITD, particularly in RV segmentation.

\textbf{Result discussions}: As one can see, in Table~\ref{tab1}, the Dice scores of UNet and TUNet using the MTS+ITD training strategy are higher than their results using the NTS+CTD training strategy. Meanwhile, the HD metrics of UNet and TUNet using the MTS+ITD training strategy are much lower than before. Despite the improvements in LV and MYO segmentation being minimal, the improvement in RV segmentation is significant. The HD metrics are reduced nearly twice compared to before. It indicates that the challenge of RV segmentation is decreased. Based on the previous analyses, the challenge of RV segmentation is mainly caused by the poor generalization ability of network models. Thus, the improvement in RV segmentation also exhibits that the generalization ability of networks is increased. Additionally, we can find that the improvement in Dice scores is less than HD metrics. We argue the reason is that the quality of original data and the construction of corresponding GT label maps. This is because the Dice scores of different networks exhibit the same score distribution, and the improvement degrees of both Dice scores are nearly the same. This improves the learning ability of networks from data has approached the limitation. Thus, the obvious improvement happens in the segmentation details, such as the HD metrics, instead of the overall segmentations, such as the Dice scores. Moreover, we have to consider the biases of data collection and annotations. In conclusion, one network model trained by different training strategies exhibits different segmentation performance. This proves that developing a more effective training strategy is significant.

\begin{figure}[t]
\newcommand{\subfigSize}{0.49}
\newcommand{\infigSize}{1.0}
\centering
	\begin{subfigure}{\subfigSize\linewidth}
     \centering
     \includegraphics[width=\infigSize\linewidth]{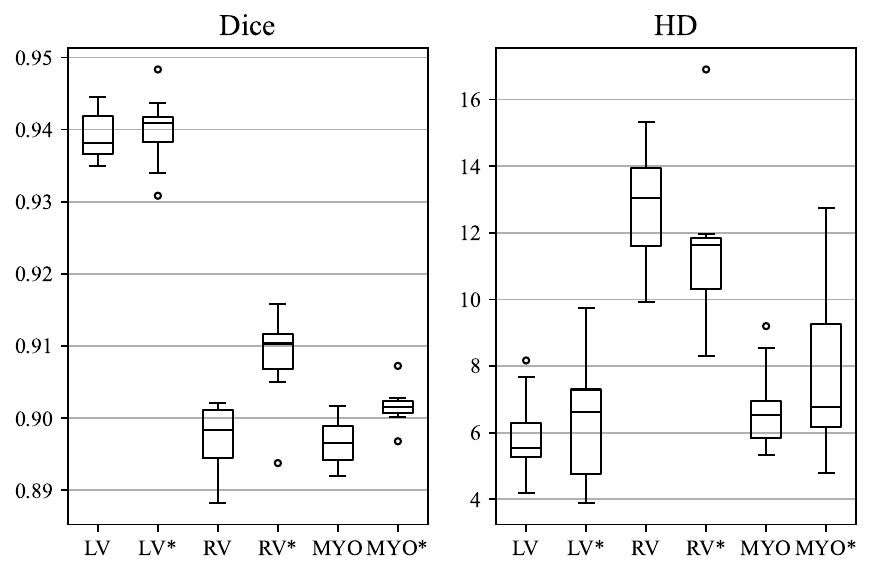}
     \caption{}
     \label{fig3a}
     \end{subfigure}
     \hfill
     \begin{subfigure}{\subfigSize\linewidth}
     \centering
     \includegraphics[width=\infigSize\linewidth]{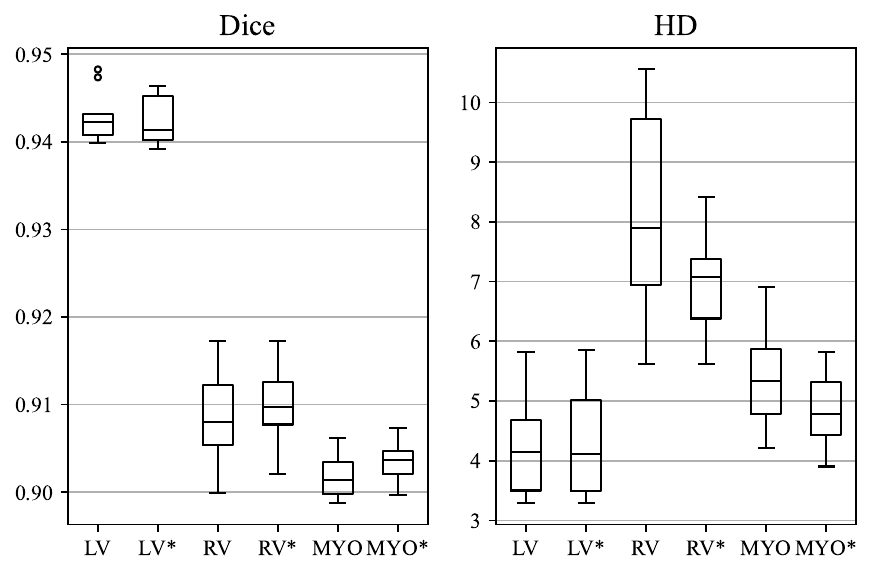}
     \caption{}
     \label{fig3b}
     \end{subfigure} 
	\vfill
	\begin{subfigure}{\subfigSize\linewidth}
     \centering
     \includegraphics[width=\infigSize\linewidth]{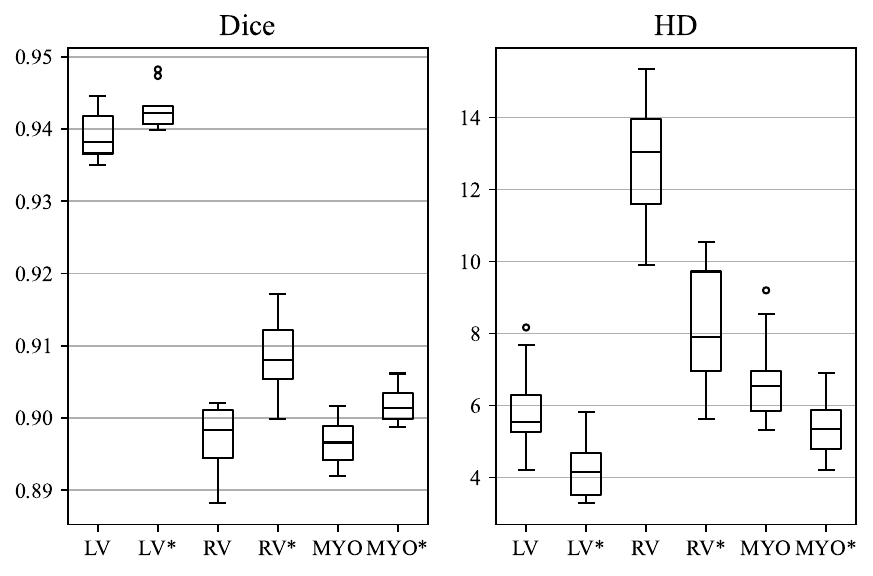}
     \caption{}
     \label{fig3c}
     \end{subfigure} 
	\hfill
	\begin{subfigure}{\subfigSize\linewidth}
     \centering
     \includegraphics[width=\infigSize\linewidth]{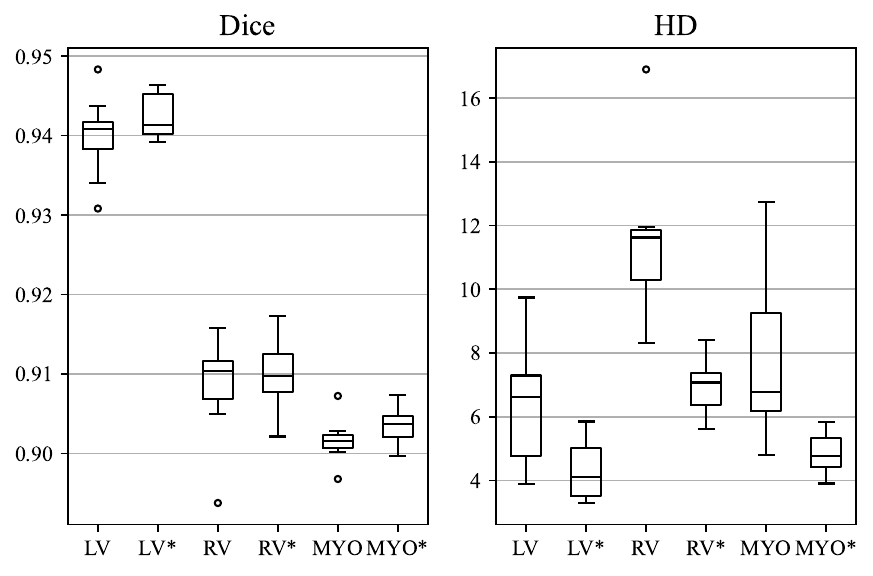}
     \caption{}
     \label{fig3d}
     \end{subfigure} 
\caption{Box plots about segmentation evaluations of UNet on ACDC testing data. (a) NTS vs. MTS* when CTD, (b) NTS vs. MTS* when ITD, (c) CTD vs. ITD* when NTS, (d) CTD vs. ITD* when MTS. Notably, the former vs. the latter, where the former lacks a star *, while the latter includes one. }
\label{fig3}
\end{figure}

\begin{figure}[t]
\newcommand{\subfigSize}{0.49}
\newcommand{\infigSize}{1.0}
\centering
	\begin{subfigure}{\subfigSize\linewidth}
     \centering
     \includegraphics[width=\infigSize\linewidth]{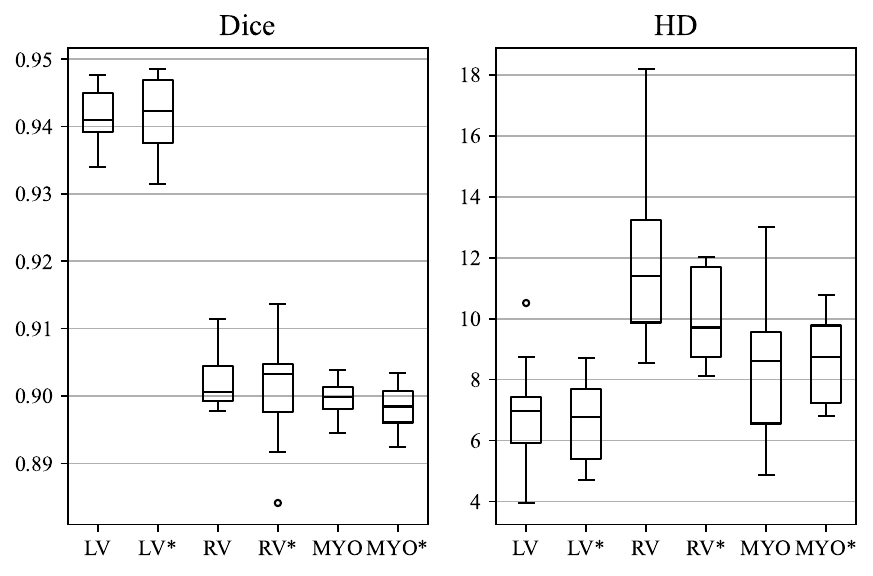}
     \caption{}
     \label{fig4a}
     \end{subfigure}
     \hfill
     \begin{subfigure}{\subfigSize\linewidth}
     \centering
     \includegraphics[width=\infigSize\linewidth]{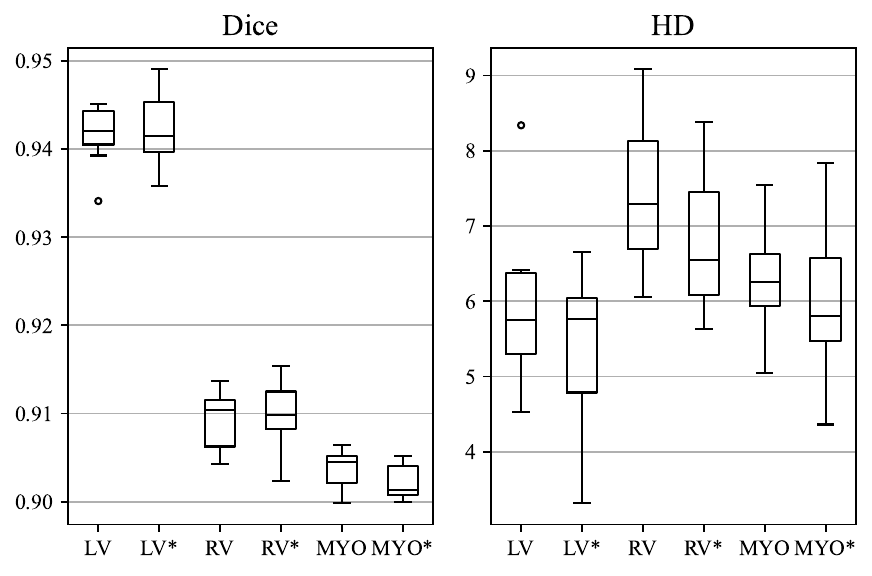}
     \caption{}
     \label{fig4b}
     \end{subfigure} 
	\vfill
	\begin{subfigure}{\subfigSize\linewidth}
     \centering
     \includegraphics[width=\infigSize\linewidth]{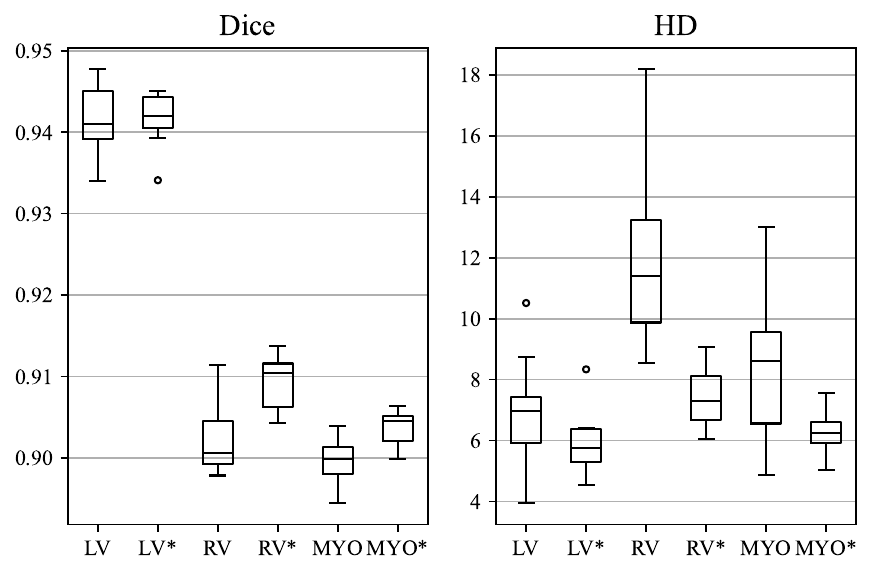}
     \caption{}
     \label{fig4c}
     \end{subfigure} 
	\hfill
	\begin{subfigure}{\subfigSize\linewidth}
     \centering
     \includegraphics[width=\infigSize\linewidth]{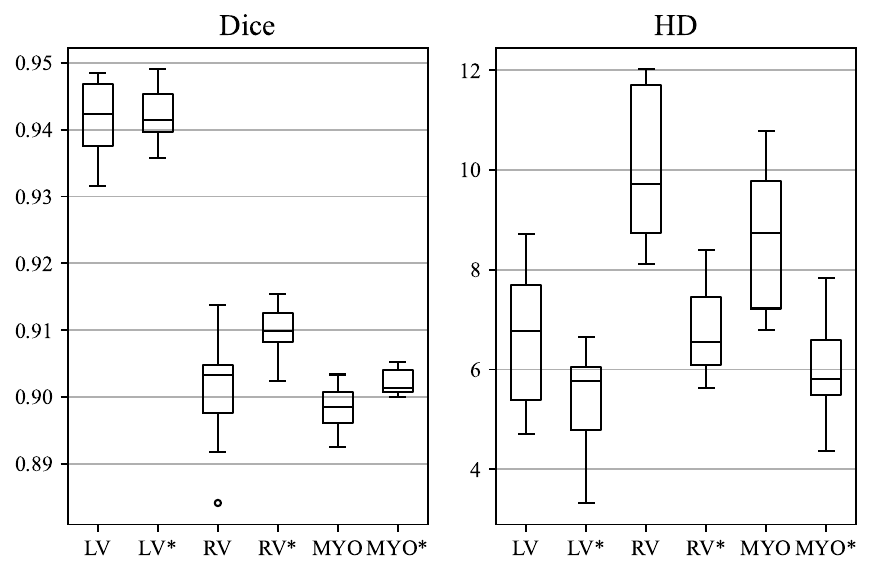}
     \caption{}
     \label{fig4d}
     \end{subfigure} 
\caption{Box plots about segmentation evaluations of TUNet on ACDC testing data. (a) NTS vs. MTS* when CTD, (b) NTS vs. MTS* when ITD, (c) CTD vs. ITD* when NTS, (d) CTD vs. ITD* when MTS. Notably, the former vs. the latter, where the former lacks a star *, while the latter includes one.}
\label{fig4}
\end{figure}

\textbf{Ablation Studies}: There are four group comparison experiments that can be regarded as ablation analyses. They are (1) NTS+CTD vs. MTS+CTD, (2) NTS+ITD vs. MTS+ITD, (3) NTS+CTD vs. NTS+ITD, and (4) MTS+CTD vs. MTS+ITD. As their name indicates, we fixed one condition and changed another to validate the effects of independent parts. The segmentation evaluations of UNet and TUNet are shown in Figures~\ref{fig3} and~\ref{fig4} respectively, and every sub-figure represents a comparison item. First, we find that MTS is superior to NTS when using the same data condition (both CTD and ITD). This is evidenced by Figures~\ref{fig3a},~\ref{fig3b},~\ref{fig4a}, and~\ref{fig4b}. Despite LV and RV evaluations being the same, the RV evaluations of the latter are better than the former in both Dice scores and HD metrics. This indicates that MTS has benefits for improving RV segmentation. Notably, MTS using Formula~\eqref{eq1} as the training objective achieved an expected solution for all disease classes. This may also result in an improvement in RV segmentation alone. Secondly, we find that ITD is superior to CTD when using the same training objective (both NTS and MTS). This is evidenced by Figures~\ref{fig3c},~\ref{fig3d},~\ref{fig4c}, and~\ref{fig4d}. All evaluations of UNet and TUNet are both improved, such as Dice scores are increased and HD metrics are decreased. This indicates that the generalization ability of networks is significantly enhanced when using incomplete training data. The occurrence of this situation is very interesting since in the training stage networks barely can see one image data twice. Therefore, if networks perform well in validation data, they need to increase their generalization ability to address any input situation. Other discussions for ITD will be shown in Section 4D. In conclusion, this ablation study provides two suggestions. They are (1) using MTS can decrease the challenge of RV segmentation and (2) incomplete training data has benefits for improving the segmentation ability and generalization ability. 

\textbf{Processing DCWC data}: This experiment purely exhibits the generalization ability since experimental data is unseen for network models. Importantly, networks are trained using ACDC data and need to process DCWC testing data directly without re-training. The segmentation evaluations of UNet and TUNet are exhibited in Table~\ref{tab2}. As we can see, both Dice scores and HD metrics are improved when using the MTS+ITD training strategy, especially in RV segmentation. This result aligns with the result in Table~\ref{tab1}, which proves that our proposed training strategy is effective both in improving RV segmentation and the generalization ability of networks. Furthermore, we can find the improvement of Dice scores is also significant, particularly for UNet models. This indicates the improvement in the generalization ability of networks is meaningful. In conclusion, according to the above experimental results, we conclude that the MTS+ITD training strategy is effective and superior when applying ACDC data. Moreover, the ability of network models to process unseen disease data is enhanced by using the MTS+ITD training strategy. 

\begin{table*}[t]
\setlength{\abovecaptionskip}{0pt}
\setlength{\belowcaptionskip}{0pt}
\caption{Cross-validation results on DCWC testing data using models trained by ACDC data.}
\centering
	\begin{subtable}{\linewidth}
	\centering
	\begin{tabular}{llrrrrrr}
	\toprule
	Training Strategies&Models&\multicolumn{2}{c}{LV Dice}&\multicolumn{2}{c}{RV Dice}&\multicolumn{2}{c}{MYO Dice}\\
	{}&{}&\multicolumn{1}{c}{ED}&\multicolumn{1}{c}{ES}&\multicolumn{1}{c}{ED}&\multicolumn{1}{c}{ES}&\multicolumn{1}{c}{ED}&\multicolumn{1}{c}{ES}\\
	\midrule
	NTS+CTD&UNet&91.3 (0.7)&85.3 (1.5)&88.4 (1.0)&85.1 (1.1)&82.3 (0.9)&84.7 (1.1)\\
	{     }&TUNet&91.9 (0.6)&86.9 (1.2)&89.2 (0.7)&85.7 (0.9)&82.9 (0.8)&85.0 (0.7)\\
	MTS+ITD&UNet&\textcolor{red}{\textbf{92.2}} (0.8)&\textbf{87.1} (1.4)&\textcolor{red}{\textbf{90.5}} (0.5)&\textcolor{red}{\textbf{86.4}} (0.4)&\textbf{82.9} (0.6)&\textbf{85.3} (0.6)\\
	{     }&TUNet&\textbf{92.2} (0.7)&\textcolor{red}{\textbf{87.2}} (0.9)&\textbf{89.9} (0.7)&\textbf{85.9} (0.6)&\textcolor{red}{\textbf{83.2}} (0.6)&\textcolor{red}{\textbf{85.4}} (0.4)\\
	\bottomrule
	\end{tabular}
	\caption{}
	\end{subtable}
	\begin{subtable}{\linewidth}
	\centering
	\begin{tabular}{llrrrrrr}
	\toprule
	Training Strategies&Models&\multicolumn{2}{c}{LV HD}&\multicolumn{2}{c}{RV HD}&\multicolumn{2}{c}{MYO HD}\\
	{}&{}&\multicolumn{1}{c}{ED}&\multicolumn{1}{c}{ES}&\multicolumn{1}{c}{ED}&\multicolumn{1}{c}{ES}&\multicolumn{1}{c}{ED}&\multicolumn{1}{c}{ES}\\
	\midrule
	NTS+CTD&UNet&12.2 (3.3)&14.8 (3.3)&13.1 (2.7)&12.5 (2.9)&13.4 (2.0)&14.5 (2.3)\\
	{     }&TUNet&11.2 (3.5)&13.2 (4.1)&12.4 (4.0)&12.3 (3.3)&12.1 (3.0)&14.9 (4.6)\\
	MTS+ITD&UNet&\textbf{8.8} (2.4)&\textcolor{red}{\textbf{10.9}} (2.1)&\textcolor{red}{\textbf{9.1}} (2.2)&\textcolor{red}{\textbf{8.6}} (1.8)&\textcolor{red}{\textbf{10.3}} (1.5)&\textcolor{red}{\textbf{10.8}} (1.7)\\
	{      }&TUNet&\textcolor{red}{\textbf{8.7}} (2.2)&\textbf{11.0} (2.0)&\textbf{9.7} (2.1)&\textbf{8.8} (1.3)&12.1 (2.0)&\textbf{13.9} (2.1)\\
	\bottomrule
	\end{tabular}
	\caption{}
	\end{subtable}
\label{tab2}
\end{table*}

\subsection{Experiments on DCWC data}
\label{sec4c}
\textbf{Experimental outline}: The experiments on DCWC data are designed to align with the experiments on ACDC data. Thus, there are still several comparison experiments that seem like the experiments in the previous section. (1) Using NTS vs. using MTS when the training data is CTD (NTS+CTD vs. MTS+CTD); (2) Using NTS vs. using MTS when the training data is ITD (NTS+ITD vs. MTS+ITD); (3) Using CTD vs. using ITD when the training strategy is NTS (NTS+CTD vs. NTS+ITD); (4) Using CTD vs. using ITD when the training strategy is MTS (MTS+CTD vs. MTS+ITD); (5) NTS+CTD vs. MTS+ITD. 

\textbf{Main segmentation results}: This experiment compares the results when using NTS and CTD training networks and the results when using MTS and ITD training networks. It \textbf{also} is the diagonal comparison in Figure~\ref{fig2}, which is NTS+CTD vs. MTS+ITD. The segmentation accuracy of processing ACDC data is shown in Table~\ref{tab3}. According to this table, we can \textbf{also} conclude two similar conclusions. (1) Different training strategies affect the segmentation ability of networks; (2) The RV segmentation ability of networks is enhanced by MTS+ITD. 

\textbf{Result discussions}: The segmentation evaluations on DCWC data are different from the results of ACDC data. As one can see, the LV and MYO segmentation evaluations are barely improved after using the MTS+ITD training strategy. In this experiment, the primary improvement only happened in RV segmentation. We argued the reasons this scenario happened are the data collection bias and the semi-professional processing technique of the GT label maps. First, the data collection bias problem we considered is the primary cause. As one can see, in the data collection stage, we collected 60 NOR cases, 30 DCM cases, 30 HCM cases, and 30 PAH cases. The number of the NOR case are twice then other disease counts. The distribution of disease in DCWC data is not the same as in ACDC data. Consequently, the segmentation evaluations are trended to the results of NOR cases. This leads to the testing segmentation results also being dominated by NOR cases, and the LV and MYO parts of NOR cases are relatively easy to partition by network models. Therefore, the LV and MYO segmentation evaluations are barely changed after using the proposed training method. Secondly, the GT label maps of DCWC data are produced in two stages: cross-dataset segmentation using one UNet model trained on the ACDC data and subsequent refinement. This UNet model is randomly selected, and the segmentation accuracy is aligned with the results of UNet models when using the NTS+CTD training strategy in Table~\ref{tab2}. Thus, we performed the refinement stage to achieve more accuracy label maps. Due to labeling 150 cases being a large workload, we split them into several clinicians. Finally, based on the personal difference, the quality of labeling is improved a little. Hence, this DCWC dataset still needs to be refined. All in all, despite several issues of DCWC data, one network model trained by different training strategies still exhibits different segmentation performance. This proves again that developing a more effective training strategy is significant.

\begin{table*}[t]
\setlength{\abovecaptionskip}{0pt}
\setlength{\belowcaptionskip}{0pt}
\caption{Segmentation evaluations on DCWC testing data.}
\centering
	\begin{subtable}{\linewidth}
	\centering
	\begin{tabular}{llrrrrrr}
	\toprule
	Training Strategies&Models&\multicolumn{2}{c}{LV Dice}&\multicolumn{2}{c}{RV Dice}&\multicolumn{2}{c}{MYO Dice}\\
	{}&{}&\multicolumn{1}{c}{ED}&\multicolumn{1}{c}{ES}&\multicolumn{1}{c}{ED}&\multicolumn{1}{c}{ES}&\multicolumn{1}{c}{ED}&\multicolumn{1}{c}{ES}\\
	\midrule
	NTS+CTD&UNet&94.0 (0.3)&\textcolor{red}{90.4} (0.3)&90.7 (0.5)&87.5 (0.5)&\textcolor{red}{85.8} (0.4)&87.6 (0.4)\\
	{     }&TUNet&93.9 (0.3)&90.3 (0.8)&90.0 (0.7)&87.0 (0.6)&85.5 (0.4)&87.4 (0.5)\\
	MTS+ITD&UNet&\textcolor{red}{\textbf{94.1}} (0.2)&90.4 (0.6)&\textcolor{red}{\textbf{90.8}} (0.7)&\textcolor{red}{\textbf{88.3}} (0.8)&85.4 (0.6)&\textcolor{red}{\textbf{87.7}} (0.5)\\
	{     }&TUNet&\textbf{94.0} (0.3)&90.3 (0.5)&\textbf{90.7} (0.6)&\textbf{88.0} (0.5)&\textbf{85.7} (0.5)&\textbf{87.6} (0.3)\\
	\bottomrule
	\end{tabular}
	\caption{}
	\end{subtable}
	\begin{subtable}{\linewidth}
	\centering
	\begin{tabular}{llrrrrrr}
	\toprule
	Training Strategies&Models&\multicolumn{2}{c}{LV HD}&\multicolumn{2}{c}{RV HD}&\multicolumn{2}{c}{MYO HD}\\
	{}&{}&\multicolumn{1}{c}{ED}&\multicolumn{1}{c}{ES}&\multicolumn{1}{c}{ED}&\multicolumn{1}{c}{ES}&\multicolumn{1}{c}{ED}&\multicolumn{1}{c}{ES}\\
	\midrule
	NTS+CTD&UNet&\textcolor{red}{3.9} (0.3)&4.8 (0.7)&10.3 (2.6)&9.7 (1.6)&\textcolor{red}{5.4} (0.2)&5.4 (0.8)\\
	{     }&TUNet&4.3 (0.5)&5.1 (1.1)&12.2 (2.3)&10.4 (2.0)&5.8 (0.6)&5.8 (0.7)\\
	MTS+ITD&UNet&4.1 (0.6)&\textcolor{red}{\textbf{4.1}} (0.8)&\textcolor{red}{\textbf{6.7}} (1.0)&\textcolor{red}{\textbf{6.7}} (1.2)&5.6 (0.7)&\textcolor{red}{\textbf{4.7}} (0.6)\\
	{      }&TUNet&4.7 (0.9)&5.4 (1.3)&\textbf{7.2} (1.0)&\textbf{7.6} (1.1)&\textbf{5.7} (0.4)&\textbf{5.5} (0.7)\\
	\bottomrule
	\end{tabular}
	\caption{}
	\end{subtable}
\label{tab3}
\end{table*}

\begin{figure}[t]
\newcommand{\subfigSize}{0.49}
\newcommand{\infigSize}{1.0}
\centering
	\begin{subfigure}{\subfigSize\linewidth}
     \centering
     \includegraphics[width=\infigSize\linewidth]{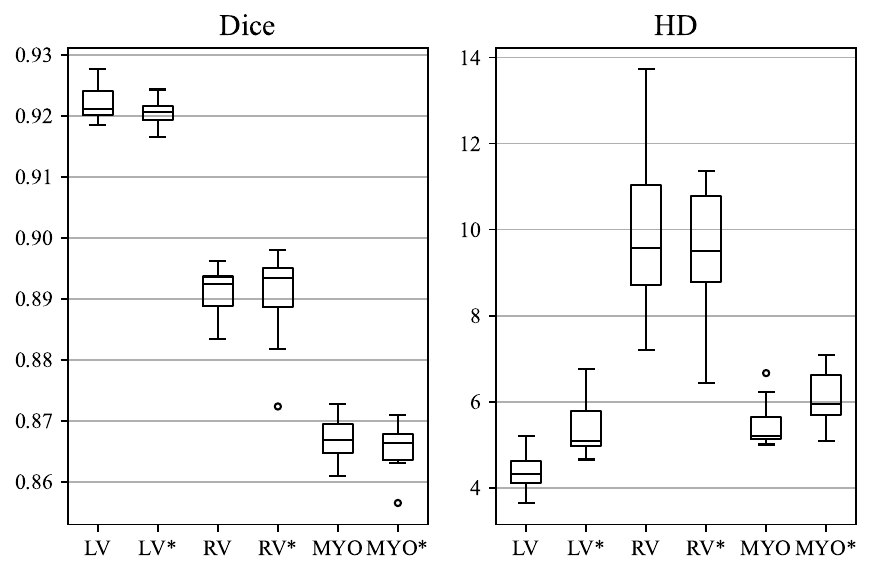}
     \caption{}
     \label{fig5a}
     \end{subfigure}
     \hfill
     \begin{subfigure}{\subfigSize\linewidth}
     \centering
     \includegraphics[width=\infigSize\linewidth]{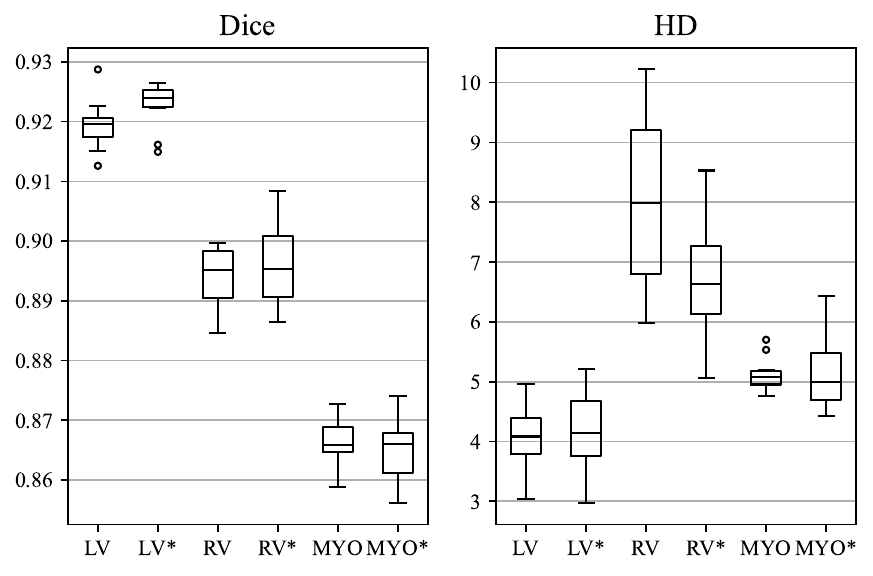}
     \caption{}
     \label{fig5b}
     \end{subfigure} 
	\vfill
	\begin{subfigure}{\subfigSize\linewidth}
     \centering
     \includegraphics[width=\infigSize\linewidth]{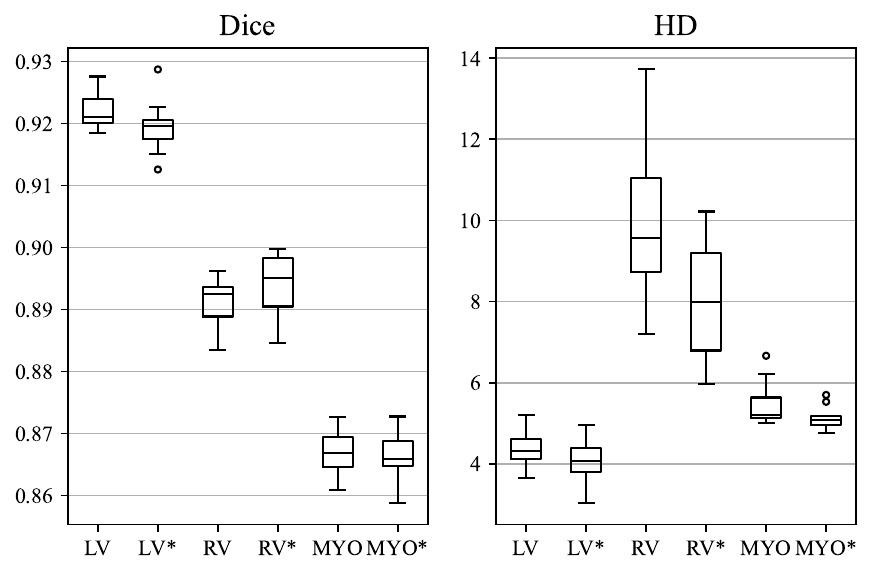}
     \caption{}
     \label{fig5c}
     \end{subfigure} 
	\hfill
	\begin{subfigure}{\subfigSize\linewidth}
     \centering
     \includegraphics[width=\infigSize\linewidth]{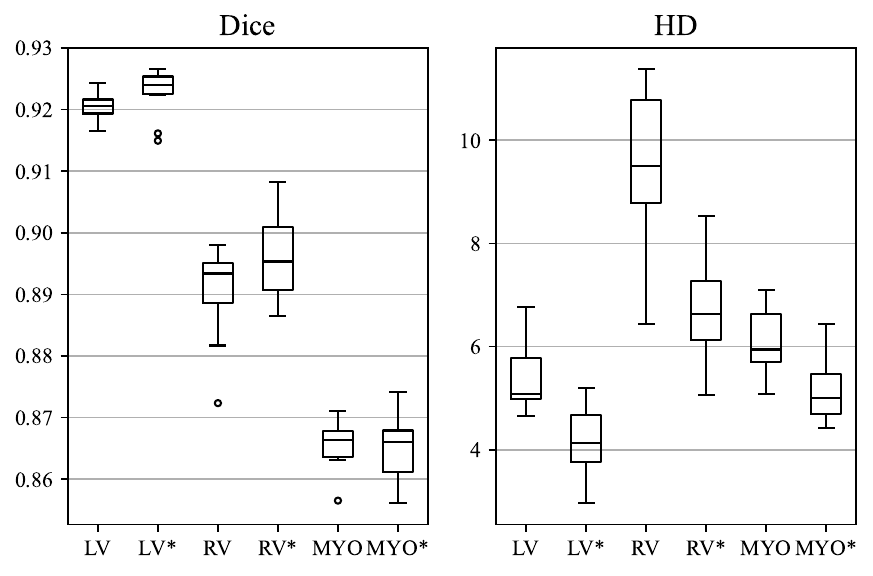}
     \caption{}
     \label{fig5d}
     \end{subfigure} 
\caption{Box plots about segmentation evaluations of UNet on DCWC testing data. (a) NTS vs. MTS* when CTD, (b) NTS vs. MTS* when ITD, (c) CTD vs. ITD* when NTS, (d) CTD vs. ITD* when MTS. Notably, the former vs. the latter, where the former lacks a star *, while the latter includes one.}
\label{fig5}
\end{figure}

\begin{figure}[t]
\newcommand{\subfigSize}{0.49}
\newcommand{\infigSize}{1.0}
\centering
	\begin{subfigure}{\subfigSize\linewidth}
     \centering
     \includegraphics[width=\infigSize\linewidth]{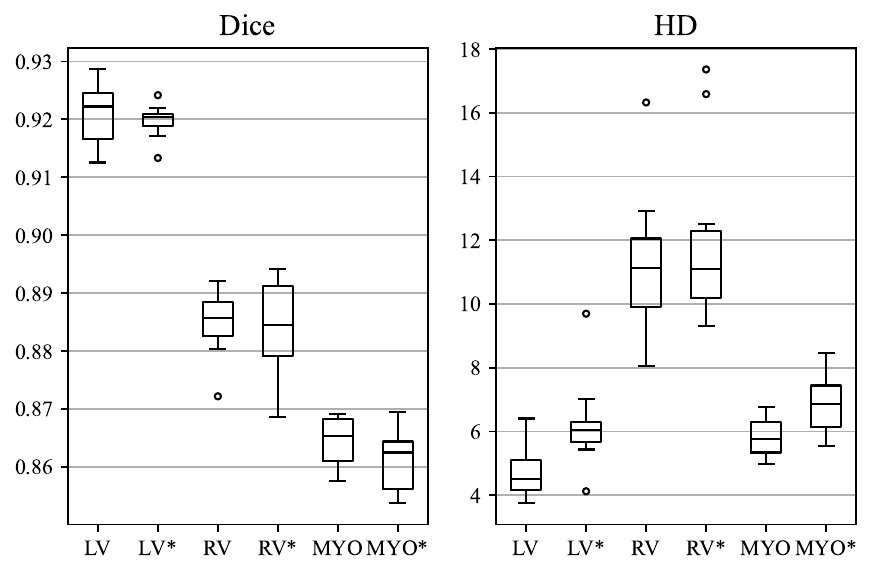}
     \caption{}
     \label{fig6a}
     \end{subfigure}
     \hfill
     \begin{subfigure}{\subfigSize\linewidth}
     \centering
     \includegraphics[width=\infigSize\linewidth]{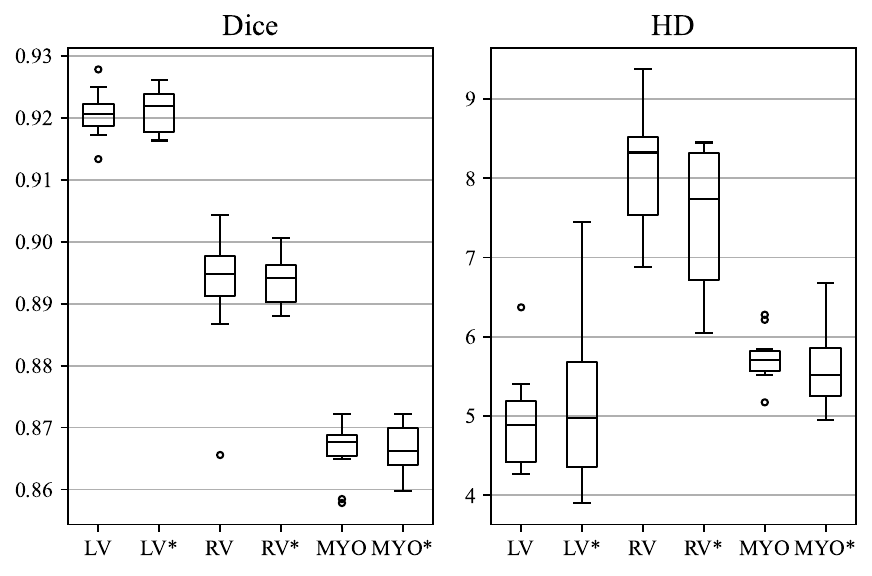}
     \caption{}
     \label{fig6b}
     \end{subfigure} 
	\vfill
	\begin{subfigure}{\subfigSize\linewidth}
     \centering
     \includegraphics[width=\infigSize\linewidth]{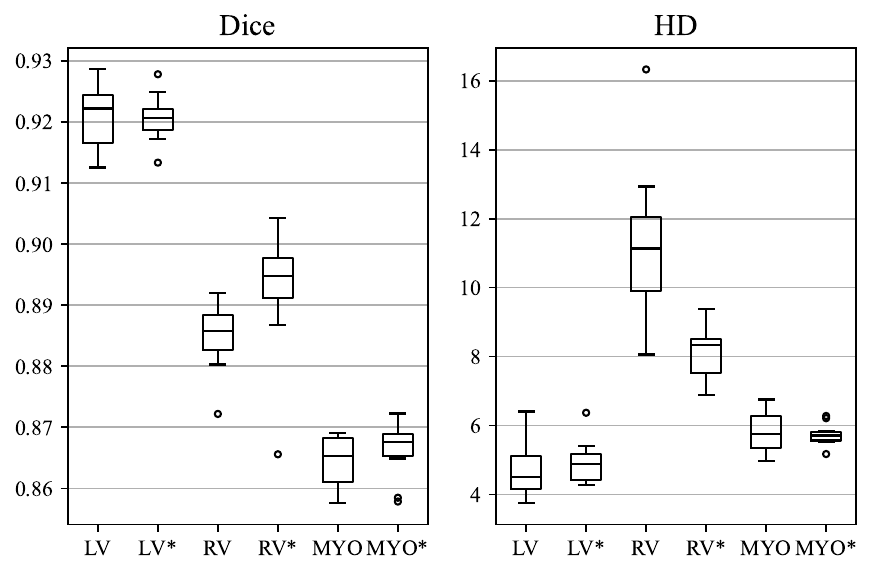}
     \caption{}
     \label{fig6c}
     \end{subfigure} 
	\hfill
	\begin{subfigure}{\subfigSize\linewidth}
     \centering
     \includegraphics[width=\infigSize\linewidth]{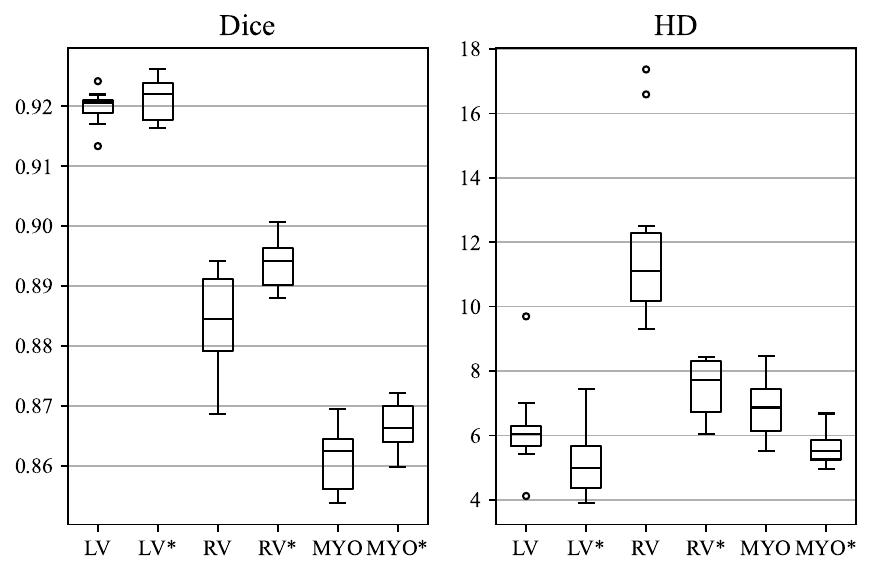}
     \caption{}
     \label{fig6d}
     \end{subfigure} 
\caption{Box plots about segmentation evaluations of TUNet on DCWC testing data. (a) NTS vs. MTS* when CTD, (b) NTS vs. MTS* when ITD, (c) CTD vs. ITD* when NTS, (d) CTD vs. ITD* when MTS. Notably, the former vs. the latter, where the former lacks a star *, while the latter includes one.}
\label{fig6}
\end{figure}

\textbf{Ablation Studies}: There are four group comparison experiments that can be regarded as ablation analyses. They are (1) NTS+CTD vs. MTS+CTD, (2) NTS+ITD vs. MTS+ITD, (3) NTS+CTD vs. NTS+ITD, and (4) MTS+CTD vs. MTS+ITD. As their name indicates, we fixed one condition and changed another to validate the effects of independent parts. The segmentation evaluations of UNet and TUNet are shown in Figures~\ref{fig5} and~\ref{fig6} respectively, and every sub-figure represents a comparison item. First, we find that the RV evaluations of the latter are better than the former in both Dice scores and HD metrics. This is evidenced by Figures~\ref{fig5a},~\ref{fig5b},~\ref{fig6a}, and~\ref{fig6b}, and this scenario is aligned with the results of ACDC data. Therefore, we argued that MTS has benefits for improving RV segmentation indeed. Secondly, we find that ITD is superior to CTD when using the same training objective (both NTS and MTS). This is evidenced by Figures~\ref{fig5c},~\ref{fig5d},~\ref{fig6c}, and~\ref{fig6d}. Despite LV evaluations being stable, other evaluations of UNet and TUNet have improved. This also can indicate that the generalization ability of networks is enhanced when using incomplete training data. The detailed discussions for ITD will be shown in Section~\ref{sec4d}, and in that section, we will analyze the affectation of the generalization ability of network models and the change of weights of network models when using ITD. In conclusion, this ablation study enhanced the conclusions in the ablation study part of the previous section.

\textbf{Processing ACDC data}: This experiment purely exhibits the generalization ability since experimental data is unseen for network models. Importantly, networks are trained using DCWC data and need to process ACDC testing data directly without re-training. The segmentation evaluations of UNet and TUNet are exhibited in Table~\ref{tab4}. As one can see, HD metrics are primarily improved when using the MTS+ITD training strategy, and this result aligns with the result in Table~\ref{tab2}. In Dice scores, the evaluations of RV are improved significantly. This proves that our proposed training strategy is effective both in improving RV segmentation and the generalization ability of networks. Furthermore, we find that the improvement in the segmentation ability of TUNet is huge, particularly the RV segmentation. This indicates that for different networks and different training data, the effectiveness of using our proposed training strategy is different. In conclusion, according to the above experimental results, we conclude that the MTS+ITD training strategy is effective and superior when applying DCWC data. Moreover, the ability of network models to process unseen disease data is enhanced by using the MTS+ITD training strategy. These conclusions and experimental results significantly validate the generality of our methods whatever data and network are. 

\begin{table*}[t]
\setlength{\abovecaptionskip}{0pt}
\setlength{\belowcaptionskip}{0pt}
\caption{Cross-validation results on ACDC testing data using models trained by DCWC data. }
\centering	
	\begin{subtable}{\linewidth}
	\centering
	\begin{tabular}{llrrrrrr}
	\toprule
	Training Strategies&Models&\multicolumn{2}{c}{LV Dice}&\multicolumn{2}{c}{RV Dice}&\multicolumn{2}{c}{MYO Dice}\\
	{}&{}&\multicolumn{1}{c}{ED}&\multicolumn{1}{c}{ES}&\multicolumn{1}{c}{ED}&\multicolumn{1}{c}{ES}&\multicolumn{1}{c}{ED}&\multicolumn{1}{c}{ES}\\
	\midrule
NTS+CTD&UNet&\textcolor{red}{94.6} (0.5)&\textcolor{red}{87.7} (0.7)&83.9 (1.2)&75.0 (1.3)&\textcolor{red}{83.6} (1.1)&\textcolor{red}{82.0} (1.6)\\
{     }&TUNet&93.2 (1.0)&85.3 (2.7)&79.9 (2.8)&70.8 (4.0)&81.7 (1.7)&79.2 (2.9)\\
MTS+ITD&UNet&94.3 (0.8)&87.0 (1.0)&\textcolor{red}{\textbf{84.8}} (3.8)&\textcolor{red}{\textbf{77.3}} (3.9)&82.8 (1.7)&81.9 (1.7)\\
{     }&TUNet&\textbf{94.3} (1.0)&\textbf{86.8} (1.5)&\textbf{84.6} (2.9)&\textbf{77.1} (3.4)&\textbf{82.8} (2.1)&\textbf{81.2} (2.6)\\
	\bottomrule
	\end{tabular}
	\caption{}
	\end{subtable}
	\begin{subtable}{\linewidth}
	\centering
	\begin{tabular}{llrrrrrr}
	\toprule
	Training Strategies&Models&\multicolumn{2}{c}{LV HD}&\multicolumn{2}{c}{RV HD}&\multicolumn{2}{c}{MYO HD}\\
	{}&{}&\multicolumn{1}{c}{ED}&\multicolumn{1}{c}{ES}&\multicolumn{1}{c}{ED}&\multicolumn{1}{c}{ES}&\multicolumn{1}{c}{ED}&\multicolumn{1}{c}{ES}\\
	\midrule
NTS+CTD&UNet&4.5 (0.6)&5.4 (0.6)&22.6 (7.9)&22.1 (5.6)&6.2 (0.6)&6.6 (0.8)\\
{     }&TUNet&4.7 (0.5)&6.6 (1.3)&26.0 (5.3)&26.2 (4.1)&6.3 (0.5)&7.7 (1.3)\\
MTS+ITD&UNet&\textcolor{red}{\textbf{3.9}} (0.7)&\textcolor{red}{\textbf{4.9}} (0.9)&\textcolor{red}{\textbf{14.8}} (2.9)&\textcolor{red}{\textbf{15.5}} (3.3)&\textcolor{red}{\textbf{5.8}} (0.3)&\textcolor{red}{\textbf{5.3}} (0.2)\\
{      }&TUNet&\textbf{4.4} (0.7)&\textbf{5.7} (1.6)&\textbf{19.2} (5.3)&\textbf{20.4} (3.7)&6.3 (0.9)&\textbf{6.4} (0.8)\\
	\bottomrule
	\end{tabular}
	\caption{}
	\end{subtable}
\label{tab4}
\end{table*}

\subsection{Analysis of ITD and Hyperparameter Experiments}
\label{sec4d}
\begin{figure}[t]
\centering
\includegraphics[width=0.50\linewidth]{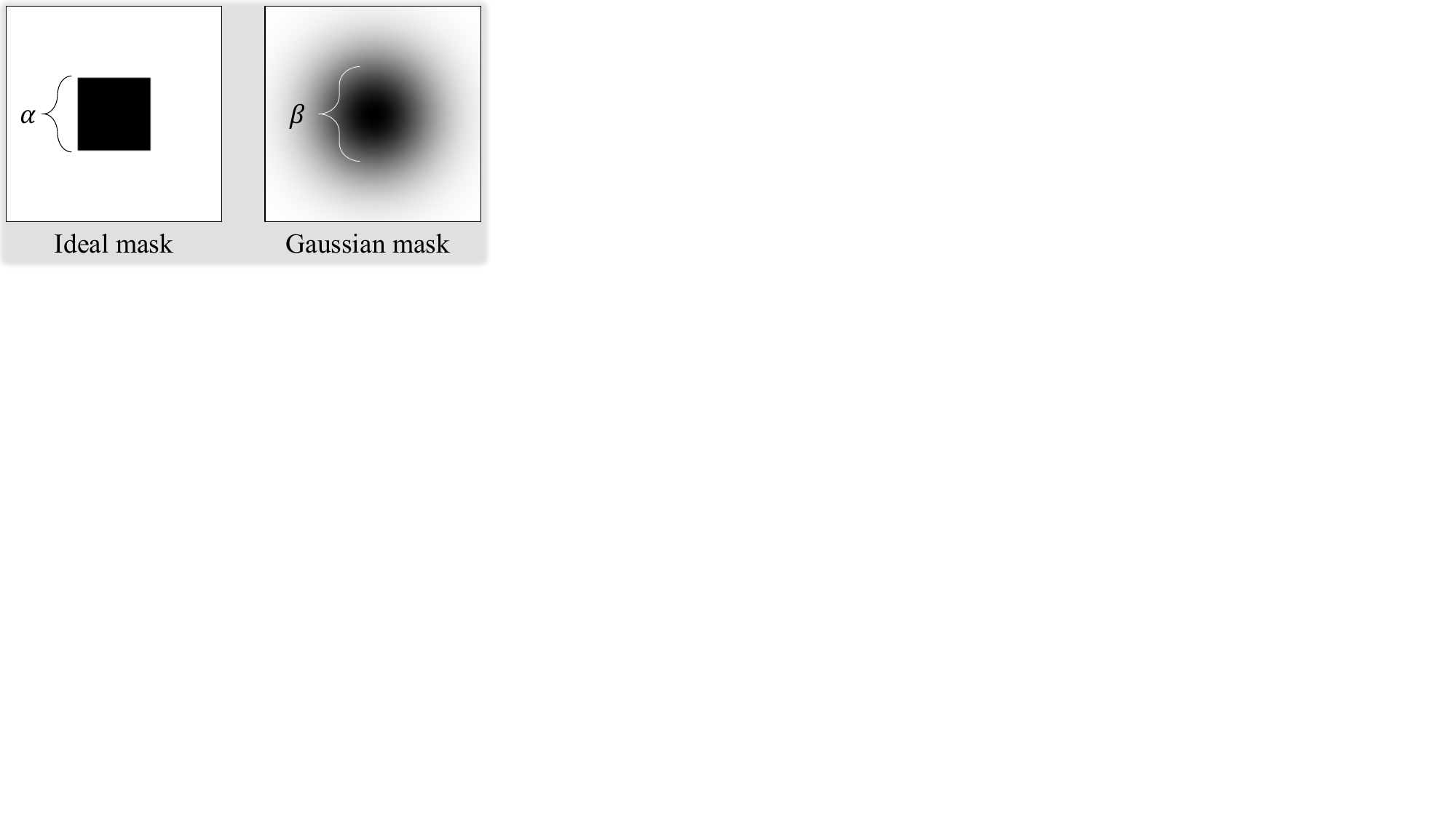}
\caption{An illustration of ideal and Gaussian masks. $\alpha$ denotes the length of a side of the ideal box, and $ \beta $ represents the diameter of the Gaussian circle.}
\label{fig7}
\end{figure}

\textbf{Ideal mask vs. Gaussian mask}: In this study, we utilize the rigidly ideal mask to destroy the part of information of training data. In contrast, we can consider a soft mask to do the same thing, such as the Gaussian mask. This mask can be formulated as 
\begin{equation}
\label{eq6}
{H_{gauss}}(u,v) = 1 - \exp ( - Dist{(u,v)^2}/2*{\beta ^2}), 
\end{equation}
where $Dist(u,v) = \sqrt[2]{{{{(u - {u_c})}^2} + {{(v - {v_c})}^2}}}$ and $ \beta  = \lambda  \cdot \min (H,W)$. Similarly, this $( {{u_c}}, {{v_c}})$ also randomly changed under a range of the image maximum length. Two different masks are shown in Figure~\ref{fig7}. As one can see, we did not utilize a circle ideal mask instead of a rectangle one. The reason is that we aim to destroy information in image data rigidly, and a circle mask like the Gaussian mask is too soft. However, a circle ideal mask is still effective, and one can conduct experiments to validate it. 

\begin{figure}[t]
\centering
\includegraphics[width=0.85\linewidth]{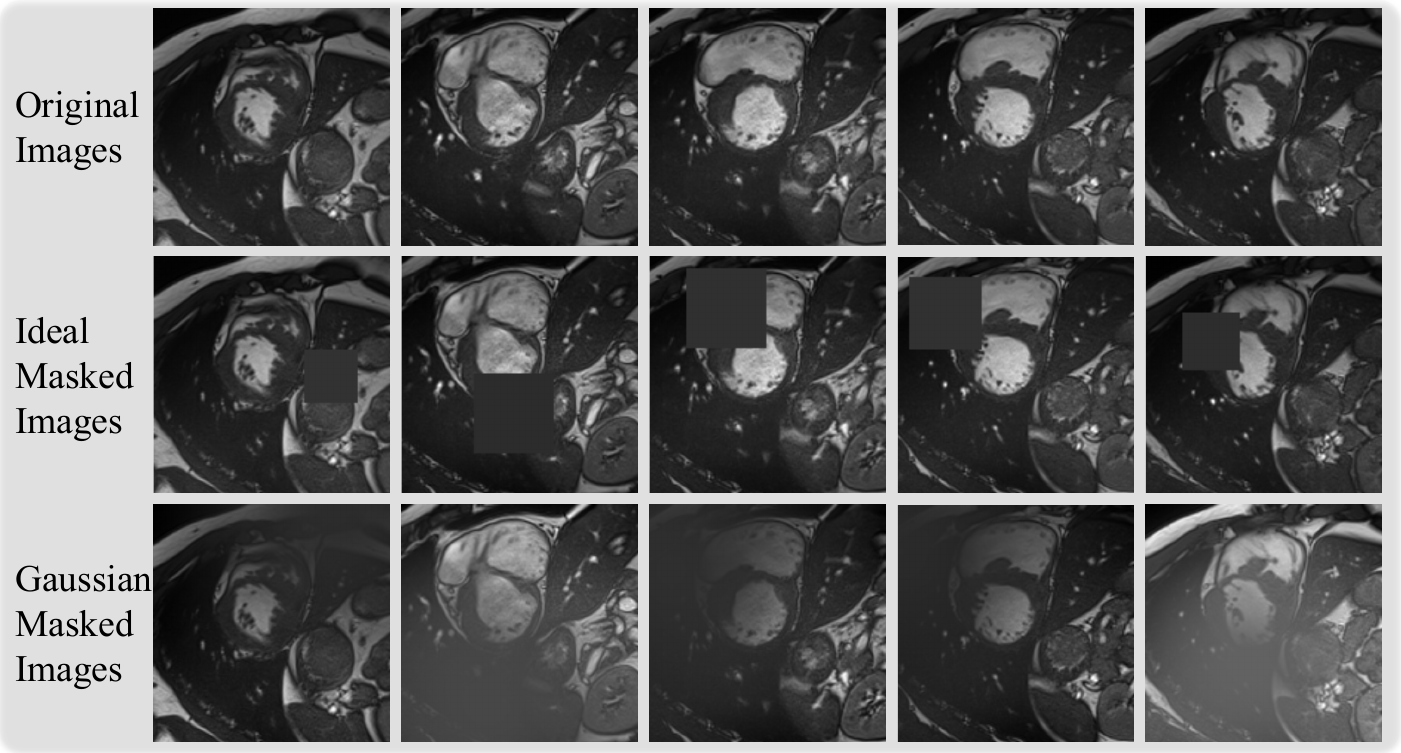}
\caption{Several examples of cardiac images before and after being preprocessed with different masks. }
\label{fig8}
\end{figure}

\begin{figure}[t]
\newcommand{\subfigSize}{0.49}
\newcommand{\infigSize}{1.0}
\centering
	\begin{subfigure}{\subfigSize\linewidth}
     \centering
     \includegraphics[width=\infigSize\linewidth]{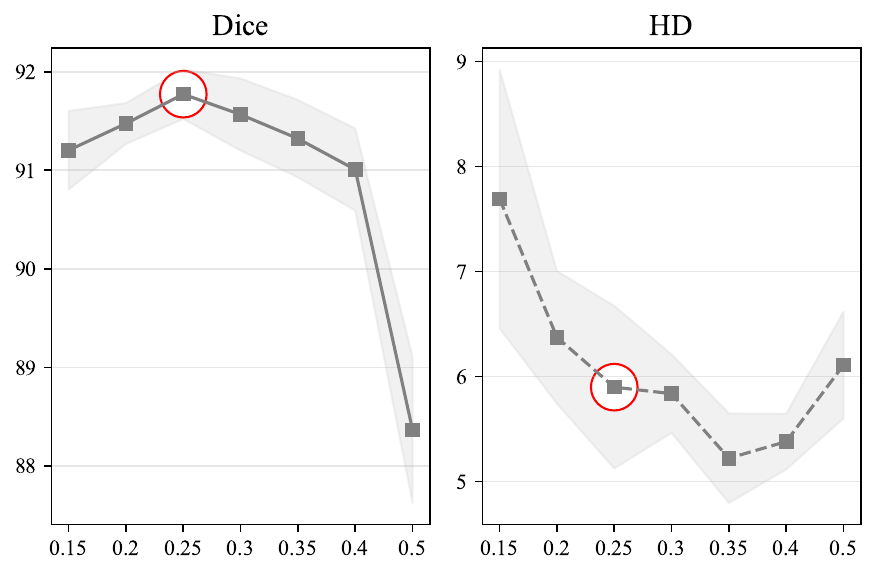}
     \caption{}
     \label{fig9a}
     \end{subfigure}
     \hfill
     \begin{subfigure}{\subfigSize\linewidth}
     \centering
     \includegraphics[width=\infigSize\linewidth]{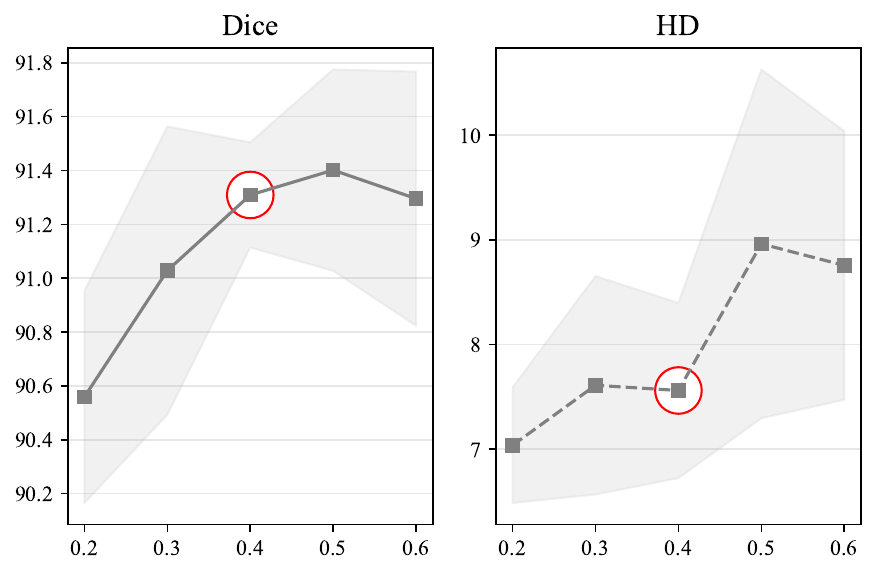}
     \caption{}
     \label{fig9b}
     \end{subfigure} 
\caption{Plots depicting the experimental results of hyper-parameter variations for both ideal and Gaussian masks. (a) shows the results for the ideal mask, while (b) displays those for the Gaussian mask. The horizontal axis represents hyperparameter values, while the vertical axis indicates Dice scores and HD metrics. Notably, the Dice scores in the plots are the average results of LV, RV, and MYO, and the HD metrics are similarly average outcomes. }
\label{fig9}
\end{figure}

The cardiac images after using the ideal mask and the Gaussian mask are shown in Figure~\ref{fig8}. The rigidness and softness are exhibited in this figure. The discarded parts in the line of ideal masked images destroyed the integrity of one image data. However, the Gaussian mask did not do that instead faded some information. To compare which mask is the most effective, we switched $H_{ideal}$ and $H_{gauss}$ in Algorithms~\ref{alg1} and \ref{alg2} and conducted the hyper-parameter $\lambda$ experiments as well as the comparison experiments. The experiment results are simply shown in Figure~\ref{fig9}, and the network is the UNet and the dataset is the ACDC dataset. As one can see, when using the ideal mask and $\lambda = 0.25 $ (see Figure~\ref{fig9a}), the Dice score is the maximum. Despite lower HD metrics existing, we let $\lambda = 0.25 $ be the final hyper-parameter of the ideal mask. Figure~\ref{fig9b} shows that when using the Gaussian mask and $\lambda = 0.4 $, the Dice score and the HD metric are relatively good. Thus, we let $\lambda = 0.4 $ be the final hyper-parameter of the Gaussian mask. Lastly, comparing Figures~\ref{fig9} and~\ref{fig9b}, we can conclude that using the ideal mask is more effective. Therefore, we find that using rigid destroyed image data to train networks can improve the generalization ability of networks, and this is the most interesting finding in this study. 

\begin{figure}[t]
\newcommand{\subfigSize}{0.49}
\newcommand{\infigSize}{1.0}
\centering
	\begin{subfigure}{\subfigSize\linewidth}
     \centering
     \includegraphics[width=\infigSize\linewidth]{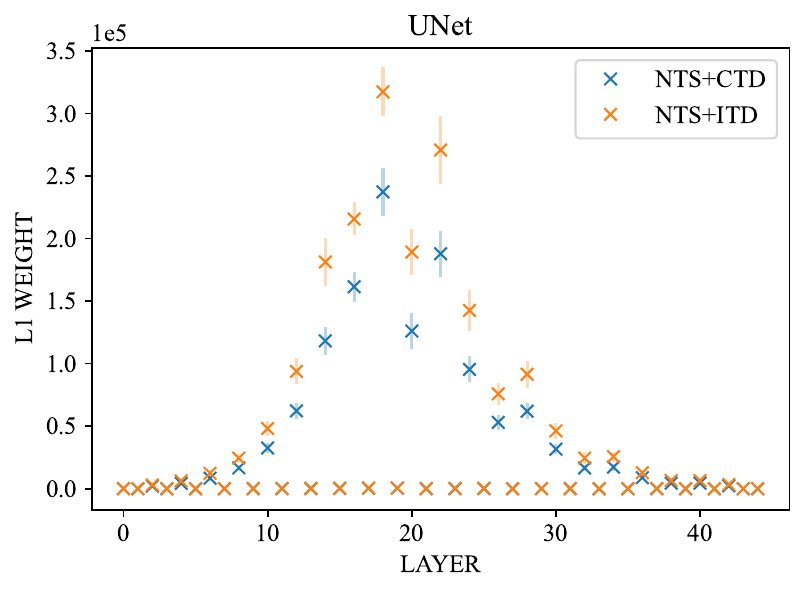}
     \caption{}
     \label{fig10a}
     \end{subfigure}
     \hfill
     \begin{subfigure}{\subfigSize\linewidth}
     \centering
     \includegraphics[width=\infigSize\linewidth]{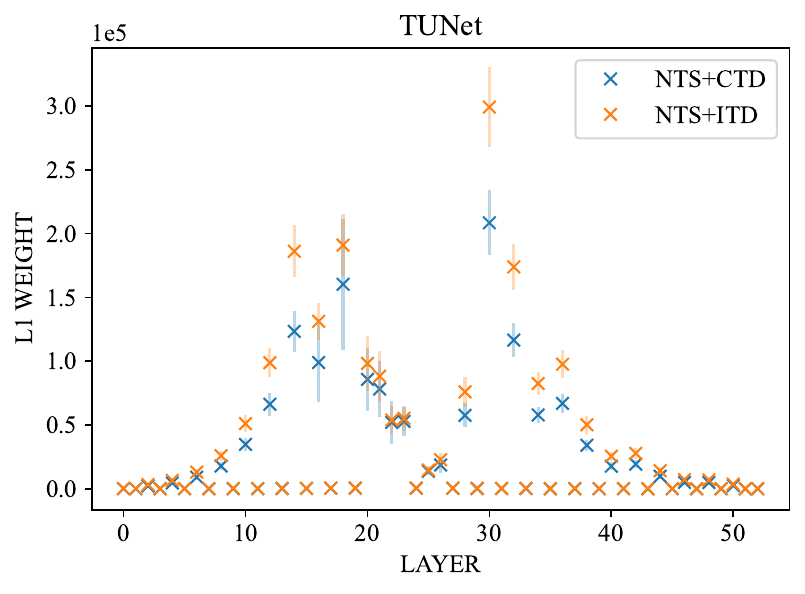}
     \caption{}
     \label{fig10b}
     \end{subfigure} 
	\vfill
	\begin{subfigure}{\subfigSize\linewidth}
     \centering
     \includegraphics[width=\infigSize\linewidth]{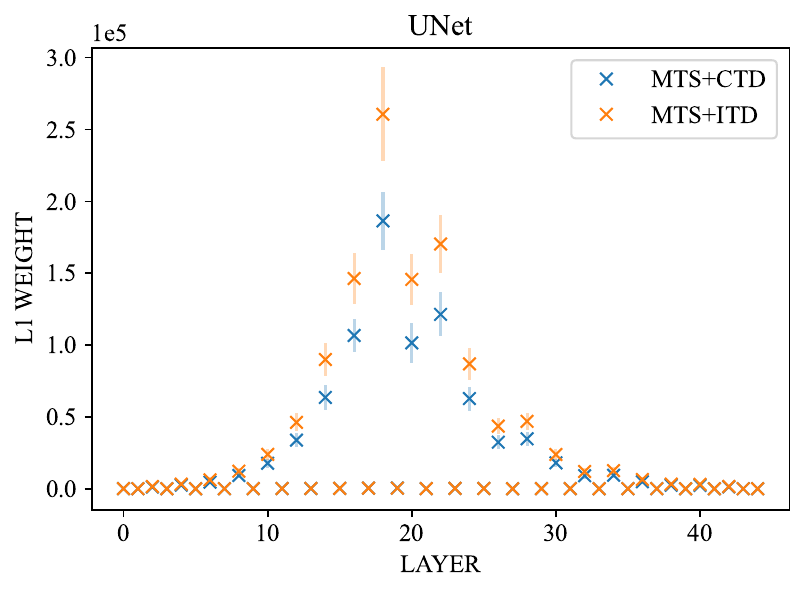}
     \caption{}
     \label{fig10c}
     \end{subfigure} 
	\hfill
	\begin{subfigure}{\subfigSize\linewidth}
     \centering
     \includegraphics[width=\infigSize\linewidth]{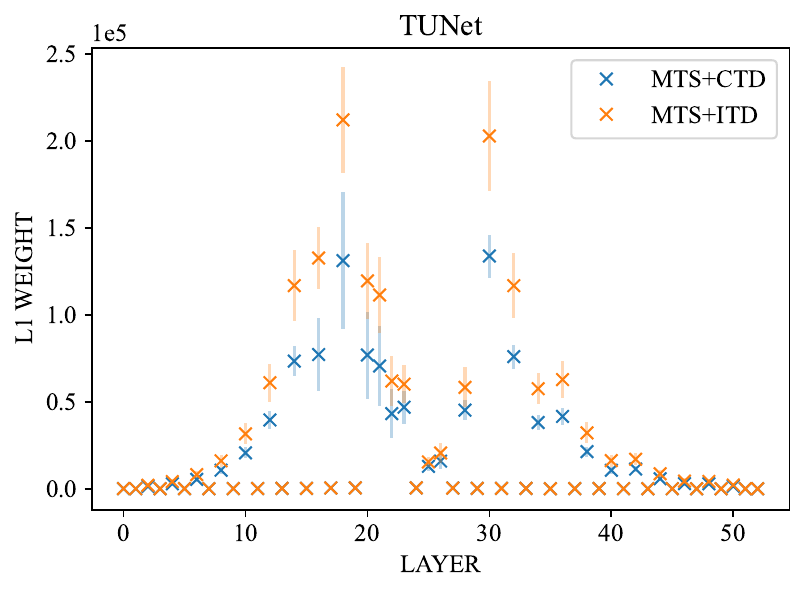}
     \caption{}
     \label{fig10d}
     \end{subfigure} 
\caption{Several plots depicting the $L_1$-norm weights of network models. (a) and (b) are UNet and TUNet when NTS, (c) and (d) are UNet and TUNet when MTS.}
\label{fig10}
\end{figure}

\textbf{Effectiveness on Ideal Mask}: After the above experiments, we are still have a confusion about why incomplete training data has such power to enhance the generalization ability of networks. Therefore, we visualized the weights of network models when using different training data conditions and different training strategies, and this visualization of the weights of networks is shown as the formation of the L1 norm. This is based on that the L1 norm of weights can directly exhibit the level of value size, and the L1 norm of weights of networks is shown in Figure~\ref{fig10}. All visualized weights are from the experiments when using ACDC data. 

In all results, we find a common phenomenon that the L1 norm of weights of networks is increased when using ITD as the training data. This is evidenced by the results of UNet and TUNet whatever the training strategy is. This scenario happened when we considered that the increased part of weights is utilized for adapting to the change of data conditions and for finding the most valuable information in incomplete inputs. It can be considered as learning causality that the primary information in input images causes the formation of the predicted label maps. As we can see, in Figure~\ref{fig8}, the ideal masked images theoretically are different even if their original images are the same. It is caused by the methods of the randomly ideal mask center selection. Therefore, in the training stage, networks barely see one image twice. However, in such cases, networks are still needed to predict the fully label maps. This enforces networks to recognize which part in inputs are contributed to the final label maps, i.e., learning the causality. Then, the weights of networks enlarged, and the generalization ability of networks increased.

\subsection{Multi-disease Segmentation Results}
\label{sec4e}
\textbf{Results on ACDC testing data}: The segmentation evaluations on five different diseases of ACDC are detailed and exhibited in Table~\ref{tab5}. According to this table, we can conclude as follows after using the MTS+ITD training strategy. (1) the segmentation results in LV and MYO of NOR and MINF are improved slightly, (2) the segmentation results in LV and MYO of DCM, HCM, and ARV are improved enormously, and (3) the most significant improvement is the RV segmentation. The first and second conclusions denote that the segmentation tasks in NOR and MINF are simpler than the tasks in other diseases, particularly the segmentation task of NOR. As we can see, in any training strategy, the segmentation results of NOR are relatively good both in Dice scores and HD metrics. This scenario aligned with the first consideration of common sense that the normal heart is easy to partition in CMRI. The third conclusion reveals the advantages of our proposed training methods. As we can see, in the RV segmentation task, Dice scores are significantly increased, and HD metrics are decreased to less than 10mm. At last, based on the data in Table~\ref{tab5}, the advantages of using our proposed methods to train network models are confirmed strongly.

\begin{table*}[t]
\setlength{\abovecaptionskip}{0pt}
\setlength{\belowcaptionskip}{0pt}
\caption{Multi-disease segmentation evaluations on ACDC testing data.}
\centering

\begin{tabular}{lllrrrrrr}
\toprule
Training Strategies & Models & Diseases & LV Dice & RV Dice & MYO Dice& LV HD & RV HD & MYO HD \\
\midrule
NTS+CTD & UNet & NOR & 93.3(5.1) & 91.4(5.9) & 90.0(2.8) & 3.2(4.2) & 7.3(6.8) & \textcolor{red}{3.7(4.4)} \\
{} & {} & MINF & 95.3(2.2) & 87.3(8.5) & 89.6(2.1) & 3.5(4.7) & 17.2(15.3) & 4.5(3.8) \\
{} & {} & DCM & 96.9(1.3) & 92.5(3.6) & 88.8(2.9) & 4.6(9.3) & 18.9(21.7) & 8.2(11.3) \\
{} & {} & HCM & 89.7(7.9) & 85.6(9.4) & 91.0(2.5) & 9.3(15.3) & 11.2(12.2) & 9.4(15.5) \\
{} & {} & ARV & 94.2(3.8) & 91.6(4.8) & 88.9(3.5) & 8.8(8.1) & 9.6(8.0) & 7.9(6.5) \\
{} & TUNet & NOR & \textcolor{red}{93.6(4.6)} & 91.7(5.3) & 90.2(2.9) & 3.1(4.8) & 6.8(8.1) & 5.1(8.8) \\
{} & {} & MINF & \textcolor{red}{95.8(1.6)} & 88.3(7.5) & 89.9(1.9) & \textcolor{red}{3.1(3.9)} & 14.6(14.8) & 4.6(4.0) \\
{} & {} & DCM & 96.7(2.2) & 92.6(3.5) & 88.7(3.0) & 7.2(14.2) & 20.1(26.6) & 9.6(14.0) \\
{} & {} & HCM & \textcolor{red}{90.6(7.2)} & 86.3(9.0) & 91.7(1.8) & 9.2(17.8) & 10.3(11.8) & 12.0(20.5) \\
{} & {} & ARV & 94.1(4.4) & 92.1(4.4) & 89.2(3.3) & 12.2(14.3) & 9.5(10.1) & 11.8(14.1) \\
\hline
MTS+ITD & UNet & NOR & 93.3(5.3) & \textcolor{red}{\textbf{92.2(5.3)}} & \textbf{90.3(2.9)} & 3.6(9.1) & \textbf{5.2(3.6)} & 4.6(6.6) \\
{} & {} & MINF & \textbf{95.8(1.7)} & \textbf{88.9(7.1)} & \textcolor{red}{\textbf{90.3(1.8)}} & 3.8(8.5) & \textbf{8.3(7.0)} & \textcolor{red}{\textbf{3.8(1.8)}} \\
{} & {} & DCM & \textcolor{red}{\textbf{97.2(1.2)}} & \textcolor{red}{\textbf{93.7(2.8)}} & \textcolor{red}{\textbf{89.4(3.0)}} & \textcolor{red}{\textbf{2.4(2.2)}} & \textbf{9.3(14.9)} & \textcolor{red}{\textbf{4.5(6.7)}} \\
{} & {} & HCM & \textbf{89.9(8.2)} & \textbf{87.9(7.9)} & \textbf{92.0(2.0)} & \textcolor{red}{\textbf{6.0(9.4)}} & \textcolor{red}{\textbf{6.0(5.2)}} & \textcolor{red}{\textbf{4.6(6.8)}} \\
{} & {} & ARV & \textcolor{red}{\textbf{95.1(3.0)}} & \textcolor{red}{\textbf{92.5(4.3)}} & \textcolor{red}{\textbf{89.8(3.0)}} & \textcolor{red}{\textbf{5.8(7.7)}} & \textcolor{red}{\textbf{6.2(3.3)}} & \textcolor{red}{\textbf{6.7(8.8)}} \\
{} & TUNet & NOR & 93.3(5.4) & \textbf{92.0(5.6)} & \textcolor{red}{\textbf{90.4(2.7)}} & \textcolor{red}{\textbf{2.7(3.5)}} & \textcolor{red}{\textbf{4.8(3.4)}} & \textbf{4.0(6.0)} \\
{} & {} & MINF & 95.6(1.8) & \textcolor{red}{\textbf{88.9(6.8)}} & \textbf{89.9(1.8)} & 3.4(7.2) & \textcolor{red}{\textbf{8.2(5.7)}} & \textbf{4.3(3.1)} \\
{} & {} & DCM & \textbf{97.2(1.2)} & \textbf{93.6(2.8)} & \textbf{89.2(2.9)} & \textbf{4.5(9.1)} & \textcolor{red}{\textbf{8.0(8.3)}} & \textbf{6.5(9.6)} \\
{} & {} & HCM & 90.3(7.4) & \textcolor{red}{\textbf{88.2(7.6)}} & \textcolor{red}{\textbf{92.1(1.7)}} & \textbf{6.2(11.2)} & \textbf{6.3(4.9)} & \textbf{5.5(10.0)} \\
{} & {} & ARV & \textbf{94.6(3.4)} & \textbf{92.4(4.2)} & \textbf{89.4(3.2)} & \textbf{10.4(11.7)} & \textbf{6.7(5.0)} & \textbf{9.8(11.8)} \\

\bottomrule
\end{tabular}

\label{tab5}
\end{table*}

\begin{table*}[t]
\setlength{\abovecaptionskip}{0pt}
\setlength{\belowcaptionskip}{0pt}
\caption{Multi-disease segmentation evaluations on DCWC testing data.}
\centering

\begin{tabular}{lllrrrrrr}
\toprule
Training Strategies & Models & Diseases & LV Dice & RV Dice & MYO Dice& LV HD & RV HD & MYO HD \\
\midrule
NTS+CTD & UNet & NOR & 93.6(5.4) & \textcolor{red}{92.1(4.9)} & 87.9(5.8) &3.2(4.0) & 5.0(5.5) & 3.9(3.5) \\
{} & {} & DCM & 90.3(7.2) & 86.1(10.7) & 84.0(7.7) & 8.3(8.8) & 11.0(10.3) & 9.2(7.4) \\
{} & {} & HCM & \textcolor{red}{95.9(2.3)} & 92.9(4.3) & 89.9(4.0) & 2.6(1.3) & 12.4(18.7) & 4.9(3.2) \\
{} & {} & PAH & \textcolor{red}{87.8(7.8)} & 82.4(12.2) & 83.8(7.6) & 4.8(5.7) & 16.8(21.8) & 5.5(4.3) \\
{} & TUNet & NOR & 93.8(4.4) & 91.7(4.9) & \textcolor{red}{87.9(5.5)} & 3.5(5.8) & 5.5(7.3) & 4.3(5.5) \\
{} & {} & DCM & 89.2(8.4) & 84.9(11.8) & 83.3(8.3) & 9.6(10.0) & 14.5(15.1) & 10.7(9.2) \\
{} & {} & HCM & 95.9(2.4) & 92.3(4.4) & \textcolor{red}{90.0(4.0)} & \textcolor{red}{2.5(1.4)} & 14.0(17.2) & \textcolor{red}{4.6(2.9)} \\
{} & {} & PAH & 87.7(8.1) & 81.9(12.8) & 83.1(8.1) & \textcolor{red}{4.6(4.4)} & 17.2(20.6) & 5.3(3.4) \\
\hline
MTS+ITD & UNet & NOR & \textcolor{red}{\textbf{93.9(4.4)}} & 92.0(4.9) & 87.6(6.0) & \textcolor{red}{\textbf{3.1(4.7)}} & \textcolor{red}{\textbf{4.3(3.2)}} & \textcolor{red}{\textbf{3.6(3.9)}} \\
{} & {} & DCM & \textcolor{red}{\textbf{90.5(8.1)}} & \textcolor{red}{\textbf{87.3(8.7)}} & \textcolor{red}{84.0(7.2)} & \textcolor{red}{\textbf{7.0(6.8)}} & \textcolor{red}{\textbf{7.7(3.3)}} & \textcolor{red}{\textbf{7.8(5.7)}} \\
{} & {} & HCM & 95.6(2.7) & 92.7(4.8) & 89.4(4.4) & 2.7(1.8) & \textcolor{red}{\textbf{7.6(9.4)}} & 5.0(3.6) \\
{} & {} & PAH & 87.6(8.5) & \textcolor{red}{\textbf{84.0(11.0)}} & \textcolor{red}{\textbf{84.0(7.3)}} & 4.9(5.7) & \textcolor{red}{\textbf{9.6(13.1)}} & 5.9(5.3) \\
{} & TUNet & NOR & 93.7(4.7) & 92.0(4.8) & 87.8(5.6) & 4.4(7.5) & \textbf{5.1(6.5)} & 4.5(5.8) \\
{} & {} & DCM & \textbf{90.0(7.3)} & \textbf{86.5(9.5)} & \textbf{83.8(7.0)} & \textbf{9.1(9.5)} & \textbf{8.7(7.3)} & \textbf{9.2(8.0)} \\
{} & {} & HCM & 95.8(2.5) & \textcolor{red}{\textbf{93.1(4.3)}} & 90.0(4.0) & 2.6(1.7) & \textbf{7.7(10.8)} & 4.9(3.3) \\
{} & {} & PAH & 87.4(8.6) & \textbf{83.2(11.9)} & \textbf{84.0(7.5)} & 5.1(5.9) & \textbf{10.7(15.2)} & \textcolor{red}{\textbf{5.2(4.1)}} \\

\bottomrule
\end{tabular}

\label{tab6}
\end{table*}

\textbf{Results on DCWC testing data}: The segmentation evaluations on four different diseases of DCWC are detailed and exhibited in Table~\ref{tab6}. According to this table, we can conclude the similar conclusions that concluded in Table~\ref{tab5}, which are as follows. (1) the segmentation results of NOR are improved slightly, and (2) the most significant improvement is the RV segmentation. These conclusions still strength the advantages of our methods. Back to Table~\ref{tab3} and the experiments on DCWC data, we analyzed why the improvement in RV segmentation is better than other organs. In Table~\ref{tab6}, we can detailly observe and analyze the segmentation situations of different diseases. The analyses of LV and MYO segmentations can be omitted, as they are easy tasks compared to the RV segmentation. Then, we only analyze the changes in RV. First, we can see that the RV HD metrics of DCM, HCM, and PAH are quite worse than NOR’s results when using the NTS+CTD training strategy. However, the RV HD metrics in Table~\ref{tab3} are relatively good due to the large number of NOR cases. The improvements of Dice scores of NOR are slight, which results in the improvements of Dice scores being tiny in Table~\ref{tab3}. This scenario ensured the analyses we argued in Section~\ref{sec4c} and proved that the data collection bias affected the segmentation evaluations of all testing data. Despite the shortages of DCWC data, we can also observe significant improvements in RV HD metrics. This improvement highlights the meaning of our proposed training strategy. Additionally, the results of DCWC reveal that our training strategy can perform well in any short-axis CMRI dataset without considering the data quality.

\section{Conclusion}
\label{sec5}
We first assume that enhancing the generalization capability of segmentation models across different slices, phases, and diseases is equivalent to improving segmentation performance for any target organ in CMRIs. Based on this assumption, we build a series of consequential works and experiments. Supported by all experimental results, we prove the effectiveness and feasibility of our proposed learning strategy. Most significantly, segmentation results on the RV improved, which is the initial motivation for our work. This proves that our assumption is reasonable and scientifically valid. Consequently, we highly encourage and recommend researchers studying medical image segmentation tasks to utilize our proposed training strategy as a basis for developing more efficient training methods.

The most interesting finding is the unexpected performance in the generalization ability of networks trained by incomplete training data. At the first insight, no one utilizes the broken training data to train networks. It is because that they consider the lost information in training data may cause network models cannot learn the important information from training data such that resulting the poor prediction ability. However, the truth is inverse. Our experimental results exhibit that destroyed data by a suitable degree has benefits for network models learning essential and crucial information from training data to increase their generalization ability. Hence, we encourage and recommend researchers to develop more effective methods for breaking training data. 

One of the main drawbacks of our approach is its heavy reliance on the quality of GT labels. This limitation is not unique to our method but rather a common challenge faced by all supervised training approaches. Additionally, we considered that achieving more accurate results requires a larger volume of data. However, this increased data requirement also leads to higher annotation costs. While transfer learning, pseudo-labels in semi-supervised learning, and synthetic data offer potential solutions, the quality of labels obtained through these methods still falls short of manual annotation. Consequently, our future research will primarily focus on developing training strategies semi-supervised or unsupervised. It will lead to a decreased demand for manual annotation and the quality of GT labels, and we are sure of that.






\bibliographystyle{IEEEtran}
\input{main.bbl}

\end{document}

%% file: main.bbl

%% file: main.bbl
\begin{thebibliography}{10}
\providecommand{\url}[1]{#1}
\csname url@samestyle\endcsname
\providecommand{\newblock}{\relax}
\providecommand{\bibinfo}[2]{#2}
\providecommand{\BIBentrySTDinterwordspacing}{\spaceskip=0pt\relax}
\providecommand{\BIBentryALTinterwordstretchfactor}{4}
\providecommand{\BIBentryALTinterwordspacing}{\spaceskip=\fontdimen2\font plus
\BIBentryALTinterwordstretchfactor\fontdimen3\font minus
  \fontdimen4\font\relax}
\providecommand{\BIBforeignlanguage}[2]{{%
\expandafter\ifx\csname l@#1\endcsname\relax
\typeout{** WARNING: IEEEtran.bst: No hyphenation pattern has been}%
\typeout{** loaded for the language `#1'. Using the pattern for}%
\typeout{** the default language instead.}%
\else
\language=\csname l@#1\endcsname
\fi
#2}}
\providecommand{\BIBdecl}{\relax}
\BIBdecl

\bibitem{b2}
P.~V. Tran, ``A fully convolutional neural network for cardiac segmentation in
  short-axis mri,'' \emph{arXiv preprint arXiv:1604.00494}, 2016.

\bibitem{b3}
O.~Ronneberger, P.~Fischer, and T.~Brox, ``U-net: Convolutional networks for
  biomedical image segmentation,'' in \emph{Medical Image Computing and
  Computer-Assisted Intervention (MICCAI 2015)}.\hskip 1em plus 0.5em minus
  0.4em\relax Springer, 2015, pp. 234--241.

\bibitem{b8}
H.~Zheng, L.~Wang, Y.~Chen, and X.~Li, ``Cross u-net: Reconstructing cardiac mr
  image for segmentation,'' in \emph{2022 IEEE International Conference on
  Multimedia and Expo (ICME)}.\hskip 1em plus 0.5em minus 0.4em\relax IEEE,
  2022, pp. 1--6.

\bibitem{b42}
K.~Zhou, Z.~Liu, Y.~Qiao, T.~Xiang, and C.~C. Loy, ``Domain generalization: A
  survey,'' \emph{IEEE Transactions on Pattern Analysis and Machine
  Intelligence}, 2022.

\bibitem{b36}
M.~Arjovsky, L.~Bottou, I.~Gulrajani, and D.~Lopez-Paz, ``Invariant risk
  minimization,'' \emph{arXiv preprint arXiv:1907.02893}, 2019.

\bibitem{b37}
M.~Choraria, I.~Ferwana, A.~Mani, and L.~R. Varshney, ``Learning optimal
  features via partial invariance,'' \emph{Proceedings of the AAAI Conference
  on Artificial Intelligence}, vol.~37, no.~6, pp. 7175--7183, Jun. 2023.

\bibitem{b12}
M.~Khened, V.~A. Kollerathu, and et~al., ``Fully convolutional multi-scale
  residual densenets for cardiac segmentation and automated cardiac diagnosis
  using ensemble of classifiers,'' \emph{Medical image analysis}, vol.~51, pp.
  21--45, 2019.

\bibitem{b13}
R.~P. Poudel, P.~Lamata, and G.~Montana, ``Recurrent fully convolutional neural
  networks for multi-slice mri cardiac segmentation,'' in \emph{Reconstruction,
  Segmentation, and Analysis of Medical Images: First International
  Workshops}.\hskip 1em plus 0.5em minus 0.4em\relax Springer, 2017, pp.
  83--94.

\bibitem{b14}
A.~Chakravarty and J.~Sivaswamy, ``Race-net: a recurrent neural network for
  biomedical image segmentation,'' \emph{IEEE journal of biomedical and health
  informatics}, vol.~23, no.~3, pp. 1151--1162, 2018.

\bibitem{b21}
A.~Ammar, O.~Bouattane, and M.~Youssfi, ``Automatic cardiac cine mri
  segmentation and heart disease classification,'' \emph{Computerized Medical
  Imaging and Graphics}, vol.~88, p. 101864, 2021.

\bibitem{b1}
J.~Long, E.~Shelhamer, and T.~Darrell, ``Fully convolutional networks for
  semantic segmentation,'' in \emph{Proceedings of the IEEE conference on
  computer vision and pattern recognition}, 2015, pp. 3431--3440.

\bibitem{b4}
{\"O}.~{\c{C}}i{\c{c}}ek, A.~Abdulkadir, and et~al., ``3d u-net: learning dense
  volumetric segmentation from sparse annotation,'' in \emph{Medical Image
  Computing and Computer-Assisted Intervention (MICCAI 2016)}.\hskip 1em plus
  0.5em minus 0.4em\relax Springer, 2016, pp. 424--432.

\bibitem{b20}
C.~Huang, H.~Han, and et~al., ``3d u 2-net: A 3d universal u-net for
  multi-domain medical image segmentation,'' in \emph{Medical Image Computing
  and Computer Assisted Intervention (MICCAI 2019)}.\hskip 1em plus 0.5em minus
  0.4em\relax Springer, 2019, pp. 291--299.

\bibitem{b15}
J.~Schlemper, O.~Oktay, and et~al., ``Attention gated networks: Learning to
  leverage salient regions in medical images,'' \emph{Medical image analysis},
  vol.~53, pp. 197--207, 2019.

\bibitem{b27}
F.~Isensee, P.~F. Jaeger, S.~A. Kohl, J.~Petersen, and K.~H. Maier-Hein,
  ``nnu-net: a self-configuring method for deep learning-based biomedical image
  segmentation,'' \emph{Nature methods}, vol.~18, no.~2, pp. 203--211, 2021.

\bibitem{b9}
A.~Vaswani, N.~Shazeer, and et~al., ``Attention is all you need,''
  \emph{Advances in neural information processing systems}, vol.~30, 2017.

\bibitem{b10}
S.~Woo, J.~Park, and et~al., ``Cbam: Convolutional block attention module,'' in
  \emph{Proceedings of the European conference on computer vision (ECCV)},
  2018, pp. 3--19.

\bibitem{b11}
S.~Song, C.~Lan, and et~al., ``An end-to-end spatio-temporal attention model
  for human action recognition from skeleton data,'' in \emph{Proceedings of
  the AAAI conference on artificial intelligence}, vol.~31, no.~1, 2017.

\bibitem{b16}
A.~Sinha and J.~Dolz, ``Multi-scale self-guided attention for medical image
  segmentation,'' \emph{IEEE journal of biomedical and health informatics},
  vol.~25, no.~1, pp. 121--130, 2020.

\bibitem{b31}
D.~Yang, A.~Myronenko, X.~Wang, Z.~Xu, H.~R. Roth, and D.~Xu, ``T-automl:
  Automated machine learning for lesion segmentation using transformers in 3d
  medical imaging,'' in \emph{Proceedings of the IEEE/CVF international
  conference on computer vision}, 2021, pp. 3962--3974.

\bibitem{b6}
O.~Oktay, E.~Ferrante, and et~al., ``Anatomically constrained neural networks
  (acnns): application to cardiac image enhancement and segmentation,''
  \emph{IEEE transactions on medical imaging}, vol.~37, no.~2, pp. 384--395,
  2017.

\bibitem{b7}
C.~Chen, C.~Biffi, and et~al., ``Learning shape priors for robust cardiac mr
  segmentation from multi-view images,'' in \emph{Medical Image Computing and
  Computer Assisted Intervention (MICCAI 2019)}.\hskip 1em plus 0.5em minus
  0.4em\relax Springer, 2019, pp. 523--531.

\bibitem{b17}
C.~Zotti, Z.~Luo, and et~al., ``Convolutional neural network with shape prior
  applied to cardiac mri segmentation,'' \emph{IEEE journal of biomedical and
  health informatics}, vol.~23, no.~3, pp. 1119--1128, 2018.

\bibitem{b18}
Q.~Zheng, H.~Delingette, and et~al., ``3-d consistent and robust segmentation
  of cardiac images by deep learning with spatial propagation,'' \emph{IEEE
  transactions on medical imaging}, vol.~37, no.~9, pp. 2137--2148, 2018.

\bibitem{b19}
Q.~Yue, X.~Luo, and et~al., ``Cardiac segmentation from lge mri using deep
  neural network incorporating shape and spatial priors,'' in \emph{Medical
  Image Computing and Computer Assisted Intervention (MICCAI 2019)}.\hskip 1em
  plus 0.5em minus 0.4em\relax Springer, 2019, pp. 559--567.

\bibitem{b25}
K.~Zhang and X.~Zhuang, ``Cyclemix: A holistic strategy for medical image
  segmentation from scribble supervision,'' in \emph{Proceedings of the
  IEEE/CVF Conference on Computer Vision and Pattern Recognition}, 2022, pp.
  11\,656--11\,665.

\bibitem{b35}
A.~Raju, S.~Miao, D.~Jin, L.~Lu, J.~Huang, and A.~P. Harrison, ``Deep implicit
  statistical shape models for 3d medical image delineation,'' in
  \emph{proceedings of the AAAI conference on artificial intelligence},
  vol.~36, no.~2, 2022, pp. 2135--2143.

\bibitem{b24}
X.~Chen, Y.~Yuan, G.~Zeng, and J.~Wang, ``Semi-supervised semantic segmentation
  with cross pseudo supervision,'' in \emph{Proceedings of the IEEE/CVF
  Conference on Computer Vision and Pattern Recognition}, 2021, pp. 2613--2622.

\bibitem{b29}
C.~M. Seibold, S.~Rei{\ss}, J.~Kleesiek, and R.~Stiefelhagen,
  ``Reference-guided pseudo-label generation for medical semantic
  segmentation,'' in \emph{Proceedings of the AAAI conference on artificial
  intelligence}, vol.~36, no.~2, 2022, pp. 2171--2179.

\bibitem{b30}
J.~Wang and T.~Lukasiewicz, ``Rethinking bayesian deep learning methods for
  semi-supervised volumetric medical image segmentation,'' in \emph{Proceedings
  of the IEEE/CVF Conference on Computer Vision and Pattern Recognition}, 2022,
  pp. 182--190.

\bibitem{b23}
X.~Yang, Y.~Zhang, B.~Lo, D.~Wu, H.~Liao, and Y.-T. Zhang, ``Dban: Adversarial
  network with multi-scale features for cardiac mri segmentation,'' \emph{IEEE
  Journal of Biomedical and Health Informatics}, vol.~25, no.~6, pp.
  2018--2028, 2020.

\bibitem{b28}
H.~Tang, X.~Liu, S.~Sun, X.~Yan, and X.~Xie, ``Recurrent mask refinement for
  few-shot medical image segmentation,'' in \emph{Proceedings of the IEEE/CVF
  international conference on computer vision}, 2021, pp. 3918--3928.

\bibitem{b41}
S.~Ben-David, J.~Blitzer, K.~Crammer, A.~Kulesza, F.~Pereira, and J.~W.
  Vaughan, ``A theory of learning from different domains,'' \emph{Machine
  learning}, vol.~79, pp. 151--175, 2010.

\bibitem{b38}
B.~Li, Y.~Shen, Y.~Wang, W.~Zhu, D.~Li, K.~Keutzer, and H.~Zhao, ``Invariant
  information bottleneck for domain generalization,'' in \emph{Proceedings of
  the AAAI Conference on Artificial Intelligence}, vol.~36, no.~7, 2022, pp.
  7399--7407.

\bibitem{b39}
H.~Zhao, C.~Dan, B.~Aragam, T.~S. Jaakkola, G.~J. Gordon, and P.~Ravikumar,
  ``Fundamental limits and tradeoffs in invariant representation learning,''
  \emph{The Journal of Machine Learning Research}, vol.~23, no.~1, pp.
  15\,356--15\,404, 2022.

\bibitem{b22}
V.~M. Campello, P.~Gkontra, C.~Izquierdo, C.~Martin-Isla, A.~Sojoudi, P.~M.
  Full, K.~Maier-Hein, Y.~Zhang, Z.~He, J.~Ma \emph{et~al.}, ``Multi-centre,
  multi-vendor and multi-disease cardiac segmentation: the m\&ms challenge,''
  \emph{IEEE Transactions on Medical Imaging}, vol.~40, no.~12, pp. 3543--3554,
  2021.

\bibitem{b26}
H.~Yao, X.~Hu, and X.~Li, ``Enhancing pseudo label quality for semi-supervised
  domain-generalized medical image segmentation,'' in \emph{Proceedings of the
  AAAI Conference on Artificial Intelligence}, vol.~36, no.~3, 2022, pp.
  3099--3107.

\bibitem{b32}
L.~Wang, D.~Li, H.~Liu, J.~Peng, L.~Tian, and Y.~Shan, ``Cross-dataset
  collaborative learning for semantic segmentation in autonomous driving,'' in
  \emph{Proceedings of the AAAI Conference on Artificial Intelligence},
  vol.~36, no.~3, 2022, pp. 2487--2494.

\bibitem{b33}
E.~Wood, T.~Baltru{\v{s}}aitis, C.~Hewitt, S.~Dziadzio, T.~J. Cashman, and
  J.~Shotton, ``Fake it till you make it: face analysis in the wild using
  synthetic data alone,'' in \emph{Proceedings of the IEEE/CVF international
  conference on computer vision}, 2021, pp. 3681--3691.

\bibitem{b34}
Y.~Ding, X.~Yu, and Y.~Yang, ``Rfnet: Region-aware fusion network for
  incomplete multi-modal brain tumor segmentation,'' in \emph{Proceedings of
  the IEEE/CVF international conference on computer vision}, 2021, pp.
  3975--3984.

\bibitem{b40}
S.~Liu, S.~Yin, L.~Qu, and M.~Wang, ``Reducing domain gap in frequency and
  spatial domain for cross-modality domain adaptation on medical image
  segmentation,'' in \emph{Proceedings of the AAAI Conference on Artificial
  Intelligence}, vol.~37, no.~2, 2023, pp. 1719--1727.

\bibitem{b5}
O.~Bernard, A.~Lalande, and et~al., ``Deep learning techniques for automatic
  mri cardiac multi-structures segmentation and diagnosis: is the problem
  solved?'' \emph{IEEE transactions on medical imaging}, vol.~37, no.~11, pp.
  2514--2525, 2018.

\end{thebibliography}
